%% file: g-paper-jstat2.tex
\documentclass[11pt]{article}
\pdfoutput=1

\usepackage{graphicx}
\usepackage[english]{babel}
\usepackage[utf8]{inputenc}
\usepackage[T1]{fontenc}
\usepackage{indentfirst}
\usepackage{amsmath}
\usepackage{amssymb}
\usepackage{eufrak}
\usepackage{psfrag}
\usepackage{pgf}

\allowhyphens

\newcommand{\complex}{\mathbb{C}}
\newcommand{\fa}{\mathfrak{a}}

\newcommand{\valos}{\mathbb{R}}
\newcommand{\eps}{\varepsilon}

\newcommand{\ordo}{\mathcal{O}}
\newcommand{\CC}{\mathcal{C}}

\newcommand{\vev}[1]{\left\langle #1 \right\rangle}
\newcommand{\ket}[1]{{\left|#1\right\rangle}}
\newcommand{\bra}[1]{{\left\langle #1\right|}}

\newcommand{\skalarszorzat}[2]{{\langle #1 | #2 \rangle}}

\newcommand{\fb}{\mathfrak{b}}
\newcommand{\fB}{\mathfrak{B}}

\def\eps{\epsilon}

\def\la{\lambda}

\def\ordo{\mathcal{O}}

\newcommand{\mc}[1]{\ensuremath{\mathcal{#1}}}
\newcommand{\mf}[1]{\ensuremath{\mathfrak{#1}}}

\newcommand\beq{\begin{equation}}
\newcommand\enq{\end{equation}}

\newcommand\bem{\begin{multline}}
\newcommand\enmu{\end{multline}}

\newcommand\beal{\begin{align}}
\newcommand\enal{\end{align}}

\def\beqa{\begin{eqnarray}}
\def\eeqa{\end{eqnarray}}
\def\ba{\begin{array}}
\def\ea{\end{array}}

\def\det{\operatorname{det}}

\def\tr{\operatorname{Tr}}

\def\Re{\operatorname{\mf{R}}}
\def\Im{\operatorname{\mf{I}}}

\newcommand{\pa}[1]{\ensuremath{\left(#1\right)}}

\newcommand{\ppa}[1]{\ensuremath{\left\{#1\right\}}}

\newcommand{\f}[2]{{\ensuremath{%
   \mathchoice%
   {\dfrac{#1}{#2}}
   {\dfrac{#1}{#2}}
   {\frac{#1}{#2}}
   {\frac{#1}{#2}}
}}}

\newcommand{\pl}[2]{\ensuremath{\prod\limits_{#1}^{#2}}}

\newcommand{\Tr}[2]{\ensuremath{\tr_{#1}\ppa{#2}}}


\usepackage{ifpdf}

\ifpdf
\usepackage{epstopdf}
\usepackage[pdftex,colorlinks,urlcolor=blue,citecolor=blue,linkcolor=blue]{hyperref}
\else
\usepackage[hypertex,colorlinks,urlcolor=blue,citecolor=blue,linkcolor=blue]{hyperref}
\fi
\begin{document}
\numberwithin{equation}{section}

\title{Exact boundary free energy of the open XXZ chain with arbitrary
  boundary conditions}
\author{B. Pozsgay$^{1,2}$ and O. Rákos$^1$\\
~\\
  $^1$ Department of Theoretical Physics, Budapest University\\
	of Technology and Economics, 1111 Budapest, Budafoki \'{u}t 8, Hungary\\
$^2$ BME Statistical Field Theory Research Group, Institute of Physics,\\
	Budapest University of Technology and Economics, H-1111 Budapest, Hungary}

\maketitle

\abstract{We derive an exact formula for the boundary free energy of the open Heisenberg XXZ
  spin chain. We allow for arbitrary boundary magnetic fields, but
  assume zero bulk magnetization. 
The result is completely analogous to
earlier formulas for the so-called $g$-function: it is expressed
 as a combination of single integrals and two
simple Fredholm determinants.
Our expressions can be evaluated easily using numerical
algorithms with arbitrary precision.
We demonstrate that the boundary free energy can show a wide variety of
behaviour as a function of the temperature, depending on the
anisotropy and the boundary fields.
We also compute the low
temperature limit of the boundary free energy, and reproduce the known
results for the ground state boundary energy, including the case of non-diagonal fields.
}

\section{Introduction}

One dimensional integrable quantum models are a special class of
many-body systems, that can be solved exactly \cite{sutherland-book}. The expression ``exact solution'' 
usually means that the eigenvalues of the Hamiltonian can be found
without approximation, and this is typically performed by one of the
forms of the Bethe Ansatz. However, integrability also offers powerful
tools to investigate physical quantities beyond the spectrum, such as
thermodynamical state functions or physical correlation functions, in
or out of equilibrium. The
calculation of such composite quantities is considerably more involved
than simply finding the spectrum.

An area that has attracted continued interest is the integrability of open
systems, and the calculation of boundary effects on the physical
quantities. The seminal work of Gaudin on the 1D Bose gas
with open boundaries \cite{Gaudin-boundary}
showed that open
systems can also be treated with the Bethe Ansatz method, and the
algebraic foundation for boundary integrability was later given in the
classic papers of Cherednik and Sklyanin
\cite{Cherednik:1985vs,sklyanin-boundary}. The key algebraic component
is the so-called Boundary Yang-Baxter (BYB) relation or ``reflection
equation'', which guarantees the commutativity of boundary transfer
matrices. This relation plays the same role as the reflection
relations in factorized scattering theory
\cite{Cherednik:1985vs,Ghoshal:1993tm}, and each solution to the BYB
equations corresponds to a specific integrable boundary condition.
In the lattice models the physically relevant solutions (those that
act as scalars on the physical Hilbert-space) are called $K$-matrices,
and they describe boundary magnetic fields. The eigenstates of such
integrable Hamiltonians are then found either by the boundary
Algebraic Bethe Ansatz \cite{sklyanin-boundary}, or by alternative methods
such as the separation of variables method
\cite{Sklyanin-sov-1,Sklyanin-sov-2} (see also
\cite{maillet-kitanine-niccoli-open-xxz-general-boundaries-1} and
references therein)  or closely
related analytical approaches \cite{off-diagonal-book}.

A paradigmatic model of one dimensional magnetism is the
$SU(2)$-symmetric Heisenberg spin chain and its anisotropic
variants. Whereas the periodic system is solved by the original Bethe Ansatz
\cite{Bethe-XXX}, the treatment of the open chain is more involved.
 In the case of longitudinal boundary magnetic fields (including free
boundary conditions) the coordinate Bethe Ansatz solution was already
given in \cite{boundaryXXZcoo}, with the algebraic formulation later
supplied in \cite{sklyanin-boundary}. The relatively simple solution
in this case made it possible to derive the boundary contributions to
the ground state energy
\cite{skorik-saleur,skorik-kapustin}.

The longitudinal fields conserve the $U(1)$-symmetry of the model,
which is reflected by the existence of a proper reference state; the
corresponding solution of the BYB equation is given
by the diagonal $K$-matrices. It was later found in
\cite{general-K-XXZ} that there is a 3-parameter family of
$K$-matrices that solve the BYB equations, and the open XXZ spin chain
is integrable with arbitrary boundary magnetic fields. The solution of
the general non-diagonal case was later given by different groups in a series of
works
\cite{nepomechie-open-1,nepomechie-open-2,kinai-open-01,offdiag1,offdiag2,niccoli-open-1,niccoli-open-2,belliard-crampe-open1,maillet-kitanine-niccoli-open-xxz-general-boundaries-1}. It
is now known that the characterization of the transfer matrix spectrum 
strongly depends on the boundary condition: the so-called T-Q
equations include an inhomogeneous term, which is zero only if the
two sets of boundary parameters satisfy a special constraint
\cite{nepomechie-open-1,nepomechie-open-2}. 
The inhomogeneous T-Q equations pose considerable challenges in the
thermodynamic limit \cite{nepomechie-elbaszott-boundary}, yet it was
possible to derive the boundary energy to the ground state using
various tricks in a series of works \cite{kinai-trukkos-boundary,kinai-su3-boundary,off-diagonal-book},
both for the massive and massless regimes of the spin chain.

Having found the characterization of the spectrum, a second step in solving a model can be the computation of its
thermodynamic properties. In integrable models the free energy density
in the infinite volume limit can be calculated by the so-called Thermodynamic
Bethe Ansatz (TBA) method \cite{yang-yang-2,zam-tba}. In the XXZ spin
chain this leads to a coupled set of non-linear integral equations (NLIE's),
where the number of the nodes depends on the anisotropy parameter
\cite{TakahashiXXX,Takahashi-Suzuki,Takahashi-book}. In the generic
case (including the isotropic point) one gets an infinite set of TBA equations.
An alternative way to describe the thermodynamics is the so-called Quantum
Transfer Matrix (QTM) method \cite{kluemper-QTM,DdV2}, which (for the
XXZ chain) leads to a single NLIE over a certain contour in the
complex plain. The QTM method is thus preferable for most problems;
its equivalence to the TBA was shown in
\cite{TBA-QTM-Kluemper-Takahashi}. 

It is a very natural idea to compute also the boundary
free energy (the $\ordo(L^0)$ contribution of a single boundary to the
free energy) using the exact methods of integrability.
The study of this quantity was initiated in the classic paper of
Affleck and Ludwig \cite{Affleck:1991tk}, where it was 
called the $g$-value (or $g$-function) or ``non-integer ground state degeneracy''. It
was conjectured in \cite{Affleck:1991tk} and later proven in \cite{Friedan:2003yc}
that the $g$-function always decreases under
renormalization group flow  and its fixed points coincide with the
$g$-values in the corresponding boundary Conformal Field Theories
(CFT's).

Early attempts to derive the $g$-function in the TBA
framework led to partial results
\cite{LeClair:1995uf,Dorey:1999cj,woynarovich}, whereas an 
exact formula was conjectured in \cite{Dorey:2004xk} based on a
low-temperature expansion in massive QFT. It was later shown in 
\cite{sajat-g,woynarovich-uj} that this exact $g$-function can be
derived by taking into account both the density of states in
rapidity space and the fluctuations around the saddle point solution
of the TBA. The general principles behind the $g$-function were thus
settled, nevertheless the result has not been applied to the XXZ spin
chains.  The formalism of \cite{sajat-g,woynarovich-uj} would lead to
Fredholm determinants that act in rapidity space and also in the infinite
dimensional string space, and their numerical or analytical evaluation
are expected to be cumbersome. 
We should note that there had been papers dealing with the boundary
free energy of the XXZ chain within the TBA approach
\cite{deSa-Tsvelik,Frahm-Zvyagin,Zvyagin-Makarova}, but they did not
include the boundary independent Fredholm determinants that were
derived only in \cite{sajat-g,woynarovich-uj}, and thus they lead to
disagreement with field theory calculations
\cite{Fujimoto:PhysRevLett92:2004,FurusakiHikihara,Affleck-XXZ-boundary-QFT}.

The computation of the boundary free energy within the QTM formalism
was initiated in \cite{Goehmann-Bortz-Frahm-boundaries},
and continued in \cite{sajat-karol}, where the $\ordo(1)$ term was
expressed as overlaps of the dominant state of the QTM with the
boundary states corresponding to the given integrable boundary
condition, in complete analogy with the picture in CFT and massive QFT
\cite{Affleck:1991tk,Dorey:2004xk}. However, the final exact result of
\cite{sajat-karol} only applied to diagonal boundary conditions, and
it was not convenient for further analytical or numerical
investigation, as it involved a complicated integral series.
The TBA results \cite{sajat-g,woynarovich-uj} together with the known parallel
between the TBA and QTM suggest
 that the boundary free energy should 
 be expressed as a combination of single integrals and simple Fredholm determinants, which
in the QTM would always be one dimensional (without string
indices).

This is the goal that we set in the present paper, and
we derive the result for generic boundary magnetic fields. Our
starting point is the formulation of the problem laid out in
\cite{Goehmann-Bortz-Frahm-boundaries,sajat-karol}, 
which we here supplement with more recent developments about the
relevant scalar products
\cite{Caux-Neel-overlap1,Caux-Neel-overlap2,sajat-minden-overlaps}.

The structure of the paper is as follows. In \ref{sec:2} we introduce
the model and our basic tools. The scalar products central to the work
are investigated in \ref{sec:overlaps}. In \ref{sec:final} we derive
the main results for the boundary free energy and the boundary
magnetization, and we also treat a number of specific cases. In
Sections \ref{sec:highT} and \ref{sec:T0} we investigate the high and
low temperature limits. Section \ref{sec:XXX} includes the formulas
specific to the XXX model, whereas in \ref{sec:ferro} we treat the
ferromagnetic XXZ spin chain. Examples for the numerical data is
presented in \ref{sec:numerics}, together with a comparison to exact
diagonalization. Finally, Section \ref{sec:conclusions} includes our
concluding remarks and the discussion of the open problems.

\section{The boundary free energy in the Quantum Transfer Matrix method}

\label{sec:2}

\subsection{Defining the model and the problem}

We define the Hamiltonian of the open anti-ferromagnetic XXZ spin chain as
\begin{equation}
  \label{H}
  \begin{split}
  H=&\sum_{j=1}^{L-1} \left\{
\sigma_j^x\sigma_{j+1}^{x}+\sigma_j^y\sigma_{j+1}^{y}+\Delta
(\sigma_j^z\sigma_{j+1}^{z}-1)
\right\}+\sum_{a=x,y,z}
h_a \left(\sigma^a_1+\sigma_L^a\right),
\end{split}
\end{equation}
where $\sigma^a$ are the Pauli matrices, and $h^a$ are the boundary
magnetic fields. For simplicity we chose to have the same magnetic fields at
the two ends. This is convenient because the contribution of a single boundary will be
given by half of the total boundary free energy. Having two different
fields would not give a different physical behaviour, because we are
taking the infinite length limit when the two ends of the chain
completely decouple.

In most of this work we restrict ourselves to
$\Delta>0$, but in Section \ref{sec:ferro} we also treat the
$\Delta<0$ regime, which (for $\Delta\le -1$) describes the ferromagnetic chain. 

In order to obtain the boundary free energy and to compare it with
previous results it is
important to fix all additive constants in the Hamiltonian. Therefore
we stress that \eqref{H} is defined such that in the absence of the
boundary fields the ferromagnetic states
with all spins up (or down) are eigenstates with zero energy. 

It is important that we did not include a bulk magnetic field in the
Hamiltonian.  The addition of a term $H'=h\sum_{j=1}^L\sigma_j^z$ is compatible
with integrability  if
 the boundary fields are also longitudinal
 ($h_x=h_y=0$). However, our current  mathematical tools
  can only be applied to the case of $h=0$; the reasons for
 this will be explained in the Conclusions. On the other
 hand, we are able to treat the case of general boundary fields, which
 is an advantage of our methods. In
 these cases $H'$ does not commute with the Hamiltonian, therefore the 
 bulk magnetic field would spoil integrability. In Conclusions we give
 further comments about this issue.

We note that the case of free boundary conditions is the
most relevant for physical applications. Apart from the treatment of
an actual open chain it can also be used to approximate
 spin chains with a finite density of non-magnetic impurities, see
 \cite{Affleck-XXZ-boundary-QFT}. Therefore, we will pay special
 attention to this case.
 
\bigskip
 
The finite temperature partition function is 
\begin{equation}
  \label{eredetiZ}
  Z\equiv \text{Tr}\ e^{-H/T}\equiv e^{-F/T}.
\end{equation}
In large volume the free energy is expected to behave as
\begin{equation}
  \label{fff}
  F=fL+2F_B+\ordo(e^{-\kappa L}),
\end{equation}
where $f$ is the free energy density defined as
\begin{equation}
  f=-\lim_{L\to\infty} \frac{T\log(Z)}{L}
\end{equation}
and the boundary contribution is
\begin{equation}
 2F_B=-\lim_{L\to \infty} \left[ T\log(Z)+fL \right].
\end{equation}
We included a factor of 2 in the definition above, such that $F_B$
describes the contribution of a single boundary. The correction terms
in \eqref{fff} decay exponentially with the volume, where $1/\kappa$
is a finite temperature correlation length.
Our goal is to compute $F_B$ in terms of the temperature and the
boundary magnetic fields.

\bigskip

The Hamiltonian \eqref{H} is integrable for any values of the boundary
magnetic fields.
This can be seen by embedding it into a family of commuting boundary
transfer matrices \cite{sklyanin-boundary}, which we briefly summarize
below.

For the construction we will use the trigonometric $R$-matrix \cite{Korepin-book} with the normalization
\begin{equation}
  \label{Rdef}
  R(u)=
  \begin{pmatrix}
    \sinh(u+\eta) & 0 & 0 & 0 \\
0 & \sinh(u) & \sinh(\eta) & 0 \\
0 & \sinh(\eta) & \sinh(u) & 0 \\
0 & 0 & 0 & \sinh(u+\eta) \\
  \end{pmatrix}.
\end{equation}
Here the parameter $\eta$ is related to the anisotropy as
$\Delta=\cosh(\eta)$; this is derived below when we relate the
Hamiltonian \eqref{H} to the transfer matrices. In the so-called massive regime with $\Delta>1$
we have $\eta\in\valos^+$, whereas in the massless case we will use
the variable $\gamma=\cos^{-1}(\Delta)$ with $\gamma\in
(0,\pi/2)$. This $R$-matrix corresponds to the statistical weights of
the 6-vertex model as given in Fig. \ref{fig:R}.

\begin{figure}[]
  \centering
\begin{pgfpicture}{0cm}{-0.2cm}{14cm}{2.2cm}
\pgfsetstartarrow{\pgfarrowcirvle{4pt}}

\pgfline{\pgfxy(0.5,1)}{\pgfxy(1.5,1)}

\pgfline{\pgfxy(1,0.5)}{\pgfxy(1,1.5)}

\pgfputat{\pgfxy(1.8,1)}{\pgfbox[center,center]{$+$}}
\pgfputat{\pgfxy(1,0.2)}{\pgfbox[center,center]{$+$}}
\pgfputat{\pgfxy(1,1.8)}{\pgfbox[center,center]{$+$}}
\pgfputat{\pgfxy(0.2,1)}{\pgfbox[center,center]{$+$}}

\pgfputat{\pgfxy(3.3,1)}{\pgfbox[center,center]{$\sinh(u+\eta)$}}

\pgfline{\pgfxy(5.6,1)}{\pgfxy(6.6,1)}  
\pgfline{\pgfxy(6.1,0.5)}{\pgfxy(6.1,1.5)}

\pgfputat{\pgfxy(6.9,1)}{\pgfbox[center,center]{$+$}}
\pgfputat{\pgfxy(6.1,0.2)}{\pgfbox[center,center]{$-$}}
\pgfputat{\pgfxy(6.1,1.8)}{\pgfbox[center,center]{$-$}}
\pgfputat{\pgfxy(5.3,1)}{\pgfbox[center,center]{$+$}}

\pgfputat{\pgfxy(7.4,1)}{\pgfbox[left,center]{$\sinh(u)$}}

\pgfline{\pgfxy(10.5,1)}{\pgfxy(11.5,1)}  
\pgfline{\pgfxy(11,0.5)}{\pgfxy(11,1.5)}

\pgfputat{\pgfxy(11.8,1)}{\pgfbox[center,center]{$-$}}
\pgfputat{\pgfxy(11,0.2)}{\pgfbox[center,center]{$-$}}
\pgfputat{\pgfxy(11,1.8)}{\pgfbox[center,center]{$+$}}
\pgfputat{\pgfxy(10.2,1)}{\pgfbox[center,center]{$+$}}

\pgfputat{\pgfxy(12.3,1)}{\pgfbox[left,center]{$\sinh(\eta)$}}
\end{pgfpicture}  
\caption{Graphical representation of the elements of the $R$-matrix.
  The rapidity $u$ is attached to the horizontal line and $0$ to the vertical
  one. Here we show 3 different non-zero matrix elements, whereas there are an
  other 3 that follow from spin reversal symmetry.
  We use the graphical convention that the $R$-matrix acts from right to left and down to up.}
  \label{fig:R}
\end{figure}

The $R$-matrix satisfies the Yang-Baxter equation
\begin{equation}
  R_{1,2}(u)R_{1,3}(u+v)R_{2,3}(v)=R_{2,3}(v)R_{1,3}(u+v)R_{1,2}(u)
\end{equation}
and the unitarity and crossing relations
\begin{equation}
  R_{1,2}(u)R_{2,1}(-u)=\sinh(u+\eta)\sinh(-u+\eta)
\end{equation}
\begin{equation}
  \label{Rcross}
  -R_{1,2}(-u)=\sigma_1^y R^{t_1}_{1,2}(u-\eta)\sigma_1^y=
  \sigma_1^y R^{t_2}_{1,2}(u-\eta)\sigma_1^y.
\end{equation}

Let $\mathcal{H}=\otimes_{j=1}^L \complex^2$ be the Hilbert space of the spin chain.
We define two monodromy matrices acting on $\mathcal{H}\otimes \complex^2$ as
\begin{equation}
  \label{ketujmonodromia_2}    
  \begin{split}
    T(\lambda)&=R_{a,L}(\lambda)\dots R_{a,1}(\lambda)\\
    \widehat{T}(\lambda)&=R_{a,1}(-\lambda)\dots R_{a,L}(-\lambda),
  \end{split}
\end{equation}
where the index $a$ refers to the auxiliary quantum space. The
boundary transfer matrix is then defined as \cite{sklyanin-boundary} 
\begin{equation}
  \label{kist}
t(\la)=\Tr{a}{K^+(\la)T(\la)K^-(\la)\widehat{T}(-\la)},
\end{equation}
where $K^\pm$ are the so-called $K$-matrices acting on
$\complex^2$. They are given as
\begin{equation}
  \label{Kpm}
 K^-(\lambda)=K(\lambda),\qquad  K^+(\lambda)=K(\lambda+\eta)
\end{equation}
with $K(\lambda)$ being the solution of the boundary Yang-Baxter (BYB)
equations \cite{Cherednik:1985vs,sklyanin-boundary}
\begin{equation}
  \label{BYB}
  R_{1,2}(u-w)K_{1}(u)R_{1,2}(u+w)K_{2}(w)=
K_{2}(w)R_{1,2}(u+w)K_{1}(u)R_{1,2}(u-w).
\end{equation}
It follows from the YB and BYB equations that for arbitrary spectral parameters
\begin{equation}
  [t(u),t(w)]=0.
\end{equation}

In the XXZ spin chain the solutions to \eqref{BYB} form a 3-parameter family \cite{general-K-XXZ}.
We use the parametrization in terms of the variables
$(\alpha,\beta,\theta)$ as
\begin{equation}
  \label{Kparam}
\begin{split}
  K_{11}(u,\alpha,\beta,\theta)=& 2(\sinh(\alpha)\cosh(\beta)\cosh(u)+ \cosh(\alpha)\sinh(\beta)\sinh(u))\\
 K_{12}(u,\alpha,\beta,\theta)=& e^\theta \sinh(2u) \\
 K_{21}(u,\alpha,\beta,\theta)=& e^{-\theta}\sinh(2u) \\
 K_{22}(u,\alpha,\beta,\theta)=& 2(\sinh(\alpha)\cosh(\beta)\cosh(u)- \cosh(\alpha)\sinh(\beta)\sinh(u)).
\end{split}
\end{equation}
The $K$-matrices satisfy the boundary crossing relation \cite{sklyanin-boundary}
\begin{equation}
  \label{Bcrossing}
  K_a(u)=\frac{1}{\sinh(2(\eta-u))}\text{Tr}_b (P_{ab}R_{ab}(-2u)K_b(u-\eta)),
\end{equation}
where $P_{ab}$ is the permutation operator.

In \eqref{Kpm} we chose the same solution of \eqref{BYB} for the two
$K$-matrices, and we will only have one set of boundary parameters
$(\alpha,\beta,\theta)$. This corresponds to having
the same boundary magnetic fields at the two ends of the chain.

The Hamiltonian is related to the first derivative of the
transfer matrix at a specific rapidity value. Although it is a textbook
exercise to establish this relation, we perform it here in
detail, in order to fix all additive constants in the Hamiltonian.

The $R$-matrix satisfies $R(0)=\sinh(\eta) P$ and
 $K^-(0)$ is proportional to the
identity, therefore
\begin{equation}
  t(0)=\frac{\sinh^{2L}(\eta)\text{Tr}(K^+(0))\text{Tr}(K^-(0))}{2}\times
  {\bf 1}.
\end{equation}
For the derivative evaluated at $u=0$ we get
\begin{equation}
  \begin{split}
 \dot t(0)=&    \sinh^{2L}(\eta)\left( \text{Tr}(K^+(0)) \dot K^-_1(0)+
   \text{Tr}(\dot K^+(0))  K^-_1(0) \right)+\\
 &+ \sinh^{2L-1}(\eta)\left[
   \sum_{j=1}^{L-1} h_{j,j+1} \text{Tr}(K^+(0)) K^-_1(0)+
   \text{Tr}(K^+_0(0) h_{0,L}) K^-_1(0)
   \right],
  \end{split}
\end{equation}
where we defined
\begin{equation}
   \label{hab}
   h_{a,b}=2P_{a,b}\dot R_{a,b}(0)=
    \sigma_a^x\sigma_{b}^{x}+\sigma_a^y\sigma_{b}^{y}+\Delta
(\sigma_a^z\sigma_{b}^{z}+1).
\end{equation}
This leads to the relation
\begin{equation}
  \label{derivrel}
     t^{-1}(0)\dot t(0)=\frac{1}{\sinh(\eta)}H_B,
\end{equation}
where $H_B$ is a  Hamiltonian operator given by
\begin{equation}
  \label{HB0}
  H_B=\sinh(\eta)\left(\frac{ \dot K^-_1(0)}{K^-_1(0)}+
    \frac{\text{Tr}(\dot K^+(0))}{\text{Tr}( K^+(0))} \right)
 +  \frac{\text{Tr}(K^+_0(0) h_{0,L})}{\text{Tr}( K^+(0))} 
+  \sum_{j=1}^{L-1}  h_{j,j+1}.
\end{equation}
It is important that $H_B$ is not identical to \eqref{H}: it differs
in certain additive terms.
Using \eqref{Kpm}, \eqref{Kparam}
and \eqref{hab}.
we obtain the explicit representation
\begin{equation}
  \label{HB}
  H_B=
  \sum_{j=1}^{L-1}  h_{j,j+1}+
\sum_{a=x,y,z}
h_a \left(\sigma^a_1+\sigma_L^a\right)
  -\frac{1}{\Delta}+2\Delta,
\end{equation}
with the boundary fields 
\begin{equation}
  \label{hdef}
  \begin{split}
    h_x&=\frac{ \sinh(\eta)  \cosh(\theta)}{\sinh(\alpha)\cosh(\beta)}\\
    h_y&=\frac{i \sinh(\eta)  \sinh(\theta)}{\sinh(\alpha)\cosh(\beta)}\\
    h_z&= \sinh(\eta)    \coth(\alpha)\tanh(\beta).
  \end{split}
\end{equation}
The parameter $\theta$ describes the angle of the
transverse magnetic fields within the $x-y$ plain. Due to rotational
symmetry we are free to set $\theta=0$, which leads to $h_y=0$.

In the massive regime $\eta\in\valos$, and real magnetic fields are
obtained if $\alpha,\beta$ are both real or if they both have imaginary
parts equal to $\pm i\pi/2$. The latter case corresponds to an
exchange of the roles of the two parameters, and for simplicity we
require that $\alpha,\beta\in\valos$. The case of purely longitudinal
fields is reached by sending one of the parameters to infinity:
\begin{equation}
  h_z=
  \begin{cases}
    \sinh(\eta)\coth(\alpha) & \text{if } \beta\to\infty\\
       \sinh(\eta)\tanh(\beta) & \text{if } \alpha\to\infty.\\
  \end{cases}
\end{equation}
Choosing the cases according to the magnitude of $h_z$ the full range can be covered with $\alpha,\beta\in\valos$.

In the massless regime $\eta=i\gamma$ with $\gamma\in\valos$, and all real
magnetic fields can be produced by $\alpha=i\tilde\alpha$ and
$\tilde\alpha,\beta\in\valos$. Longitudinal fields are reached when
$\beta\to\infty$ such that
\begin{equation}
  h_z=\sin(\gamma)\cot(\tilde\alpha).
\end{equation}
Free boundary conditions are obtained further by setting $\tilde\alpha=\pi/2$.

For practical purposes we also present the inversion of the relations
\eqref{hdef} (with $h_y=0$):
\begin{equation}
  \label{albefromh}
  \begin{split}
    \alpha&=\frac{1}{2}\left[
      \sinh^{-1}\left(\frac{\sinh(\eta)+h_z}{h_x}\right)+
      \sinh^{-1}\left(\frac{\sinh(\eta)-h_z}{h_x}\right)    \right]\\
\beta&=\frac{1}{2}\left[
  \sinh^{-1}\left( \frac{\sinh(\eta)+h_z}{h_x}\right)
  -\sinh^{-1}\left(  \frac{\sinh(\eta)-h_z}{h_x}\right)  \right].
    \end{split}
\end{equation}
These relations are valid in both regimes and they produce the
$\alpha,\beta$ in accordance with the previous discussion.

In \eqref{HB} the two boundary terms are exactly equal, even
though they originate from formally different expressions in
\eqref{HB0}. This is not a coincidence: their agreement can be
proven from the crossing relation \eqref{Bcrossing}, as it was
originally remarked by Sklyanin \cite{sklyanin-boundary}.

The differences in the additive terms in $H$ and $H_B$
originate from  the overall normalization of the $R$-matrix and the
$K$-matrices.
In the intermediate calculations we will work with $H_B$, because certain
 expressions are simpler this way. The original Hamiltonian is then
restored through the relation
\begin{equation}
  \label{Hrel}
  H=H_B-2\Delta L +\frac{1}{\Delta}.
\end{equation}
This intermediate relation is singular
for the specific $\Delta\to 0$ limit, but our final formulas for the boundary free energy
(with respect to $H$) have a finite $\Delta\to 0$ limit.

\subsection{The TBA approach}

\label{sec:TBA}

A natural way to evaluate thermal partition functions in integrability
is to express them as a sum over all eigenstates of the system, to derive
the Bethe equations characterizing the spectrum, and to perform a
saddle point approximation of the summation to obtain the free energy. This method is known
as the Thermodynamic Bethe Ansatz (TBA)
\cite{yang-yang-2,Takahashi-book}. Originally devised to compute the
free energy density, the TBA is also capable to capture all  $\ordo(1)$
contributions to the free energy: It was shown in
\cite{sajat-g,woynarovich-uj}, that the resulting finite terms include simple
integrals (that arise from rapidity shifts of the bulk particles due
to the boundaries) and two boundary-independent Fredholm determinants,
which result from the fluctuations around the saddle point solution of
the TBA, and the non-trivial density of states in rapidity space. 
The papers \cite{sajat-g,woynarovich-uj} provided very general results,
and in principle they could be applied to the XXZ spin chain as
well. However, there are two difficulties associated with this approach.

First of all, the Bethe Ansatz equations of the boundary spin chain do not always
take the usual product form that was assumed in
\cite{sajat-g,woynarovich-uj}. In fact, for generic boundary
parameters one has to deal with inhomogeneous T-Q equations and
polynomial equations describing the spectrum
\cite{kinai-open-01,offdiag1,offdiag2,niccoli-open-1,niccoli-open-2,belliard-crampe-open1,maillet-kitanine-niccoli-open-xxz-general-boundaries-1}. In
principle this difficulty could be avoided by concentrating on those
special cases \cite{nepomechie-open-1,nepomechie-open-2,off-diagonal-book} where the
inhomogeneous terms vanish and the usual Bethe Ansatz equations are
recovered. In the thermodynamic limit these special points become dense in the
space of boundary conditions, and by continuity they would give the
boundary free energy for arbitrary magnetic fields \cite{off-diagonal-book}. 

A second difficulty with the TBA approach is that the resulting Fredholm
determinants reflect the particle content of the theory, and they act
in the direct sum of the rapidity spaces of all the particles. 
The XXZ spin chain typically has an infinite number of string
solutions, and in the TBA they are treated as independent
particles. This would lead to infinite dimensional Fredholm
determinants\footnote{These Fredholm determinants
were actually presented in \cite{sajat-minden-overlaps}, although in
a different context, see subsection 2.5. there.}.

We believe that the TBA would lead to correct final results, but their
numerical implementation would be demanding.
In the present work we take a different approach. We compute the boundary free
energy using the Quantum Transfer Matrix method, and show
that this leads to single
integrals and one dimensional Fredholm determinants. In the
Conclusions we give a few additional remarks about the relation
between the two approaches.

The QTM method does not require the knowledge of the spectrum of the
original Hamiltonian \eqref{H}, therefore we do not review its
solution. We refer the interested reader to the papers cited above. 

\subsection{The QTM approach}

Following \cite{Goehmann-Bortz-Frahm-boundaries} we develop a lattice
path integral for the partition
function \eqref{eredetiZ}, and we evaluate it in the large volume
limit. The key idea is to build an alternative transfer matrix (called
the QTM) which acts along the space direction of the corresponding 2D
statistical physical system \cite{kluemper-QTM}. This is completely
analogous with the usual step of exchanging the space
and time directions in integrable (euclidean) QFT \cite{zam-tba,Dorey:2004xk}.

It follows from \eqref{derivrel} that exponentials of $H_B$ can be evaluated in the so-called Trotter
expansion as
\begin{equation}
\text{Tr}\ e^{-H_B/T}  =\lim_{N\to\infty} \text{Tr}\left(1-\frac{H_B}{TN}\right)^N=
\lim_{N\to\infty} t^{-N}(0)\bar Z(N,L),
\end{equation}
where we defined
\begin{equation}
\label{Zbar}
\bar Z(N,L)=
\text{Tr}\ t^N(-\omega).
\end{equation}
Here $N$ is called the Trotter number, and $\omega$ is an
$N$-dependent parameter given by
\begin{equation}
  \label{omegadef}
 \omega=\frac{\sinh\eta}{NT}.
\end{equation}
Certain intermediate formulas in the following sections can depend
on the parity of $N$. For simplicity we will assume that $N$ is even.

Comparing to \eqref{Hrel} we find that
\begin{equation}
  \label{Z1}
  Z=e^{-F/T}=
e^{-\frac{1}{\Delta T}+\frac{2\Delta L}{T}}  \lim_{N\to\infty} \Big[ t^{-N}(0)\bar Z(N,L)\Big].
\end{equation}

\begin{figure}
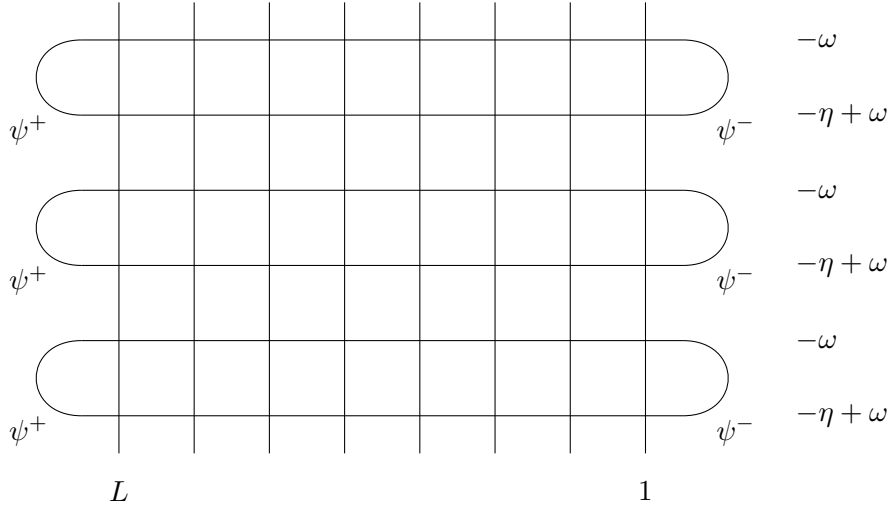

\centering
\begin{pgfpicture}{0cm}{0cm}{12cm}{7cm}

\pgfsetendarrow{\pgfarrowto}

\pgfline{\pgfxy(1.5,1)}{\pgfxy(9.5,1)}
\pgfline{\pgfxy(1.5,2)}{\pgfxy(9.5,2)}
\pgfline{\pgfxy(1.5,3)}{\pgfxy(9.5,3)}
\pgfline{\pgfxy(1.5,4)}{\pgfxy(9.5,4)}
\pgfline{\pgfxy(1.5,5)}{\pgfxy(9.5,5)}
\pgfline{\pgfxy(1.5,6)}{\pgfxy(9.5,6)}

\pgfputat{\pgfxy(11,2)}{\pgfbox[left,center]{$-\omega$}}
\pgfputat{\pgfxy(11,4)}{\pgfbox[left,center]{$-\omega$}}
\pgfputat{\pgfxy(11,6)}{\pgfbox[left,center]{$-\omega$}}
\pgfputat{\pgfxy(11,1)}{\pgfbox[left,center]{$-\eta+\omega$}}
\pgfputat{\pgfxy(11,3)}{\pgfbox[left,center]{$-\eta+\omega$}}
\pgfputat{\pgfxy(11,5)}{\pgfbox[left,center]{$-\eta+\omega$}}

\pgfputat{\pgfxy(9,0)}{\pgfbox[center,center]{$1$}}
\pgfputat{\pgfxy(2,0)}{\pgfbox[center,center]{$L$}}

\pgfputat{\pgfxy(10.2,0.8)}{\pgfbox[center,center]{$\psi^-$}}
\pgfputat{\pgfxy(10.2,2.8)}{\pgfbox[center,center]{$\psi^-$}}
\pgfputat{\pgfxy(10.2,4.8)}{\pgfbox[center,center]{$\psi^-$}}
\pgfputat{\pgfxy(0.8,0.8)}{\pgfbox[center,center]{$\psi^+$}}
\pgfputat{\pgfxy(0.8,2.8)}{\pgfbox[center,center]{$\psi^+$}}
\pgfputat{\pgfxy(0.8,4.8)}{\pgfbox[center,center]{$\psi^+$}}

\pgfline{\pgfxy(2,0.5)}{\pgfxy(2,6.5)}
\pgfline{\pgfxy(3,0.5)}{\pgfxy(3,6.5)}
\pgfline{\pgfxy(4,0.5)}{\pgfxy(4,6.5)}
\pgfline{\pgfxy(5,0.5)}{\pgfxy(5,6.5)}
\pgfline{\pgfxy(6,0.5)}{\pgfxy(6,6.5)}
\pgfline{\pgfxy(7,0.5)}{\pgfxy(7,6.5)}
\pgfline{\pgfxy(8,0.5)}{\pgfxy(8,6.5)}
\pgfline{\pgfxy(9,0.5)}{\pgfxy(9,6.5)}

\pgfclearendarrow

\pgfmoveto{\pgfxy(9.5,5)}
\pgfcurveto{\pgfxy(10.3,5)}{\pgfxy(10.3,6)}{\pgfxy(9.5,6)}
\pgfstroke
\pgfmoveto{\pgfxy(9.5,3)}
\pgfcurveto{\pgfxy(10.3,3)}{\pgfxy(10.3,4)}{\pgfxy(9.5,4)}
\pgfstroke
\pgfmoveto{\pgfxy(9.5,1)}
\pgfcurveto{\pgfxy(10.3,1)}{\pgfxy(10.3,2)}{\pgfxy(9.5,2)}
\pgfstroke

\pgfmoveto{\pgfxy(1.5,5)}
\pgfcurveto{\pgfxy(0.7,5)}{\pgfxy(0.7,6)}{\pgfxy(1.5,6)}
\pgfstroke
\pgfmoveto{\pgfxy(1.5,3)}
\pgfcurveto{\pgfxy(0.7,3)}{\pgfxy(0.7,4)}{\pgfxy(1.5,4)}
\pgfstroke
\pgfmoveto{\pgfxy(1.5,1)}
\pgfcurveto{\pgfxy(0.7,1)}{\pgfxy(0.7,2)}{\pgfxy(1.5,2)}
\pgfstroke

\end{pgfpicture}
\caption{Lattice path integral representation for the partition
  function $\bar Z(N,L)$. The vertices are given by the six-vertex
  model as depicted in Fig \ref{fig:R}, the boundary weights on the
  left and the right are
  are related to the $K$-matrices (see eqs. \eqref{vectors}-\eqref{vectors2}) and there are
  periodic boundary conditions in the time directions. }
\label{startingpoint2}
\end{figure}

With the repeated use of \eqref{Rcross} one gets
\begin{equation}
  \label{Tcross}
\hat  T(-u)=\sigma_0^y T^{t_0}(-u-\eta)\sigma_0^y.
\end{equation}
It follows that $\bar Z(N,L)$
is equal to the partition function of the six vertex model with
boundaries and spectral parameters as specified in the figure
\ref{startingpoint2}, where the vertices are given by the $R$-matrix
\eqref{Rdef} and the boundary weights follow from the $K$-matrices;
explicit formulas will be given below.

The main idea to evaluate this partition function in the large volume
limit is to build a transfer matrix that acts in the horizontal
direction. This operator is called the
Quantum Transfer Matrix \cite{kluemper-QTM,kluemper-review}. It can be
established algebraically by a reordering of the $R$-matrices under
the trace in \eqref{Zbar} (for a detailed computation relevant to the present problem see
\cite{sajat-karol}). Alternatively, it can be constructed by
performing a reflection of the partition function of Fig.
\ref{startingpoint2} around the North-East axis, and by building a new
transfer matrix that acts in the vertical direction, but acting on a
spin chain with alternating inhomogeneities. This leads to the
construction of the ``quantum monodromy matrix'' as
\begin{equation}
  \label{TQTM}
  T_{QTM}(u)=R_{2N,0}(u-\eta+\omega)R_{2N-1,0}(u-\omega)
  \dots R_{2,0}(u-\eta+\omega)R_{1,0}(u-\omega),
\end{equation}
where $\omega$ is the parameter defined in \eqref{omegadef}.
The Quantum Transfer Matrix is then given by
\begin{equation}
  t_{QTM}(u)=\text{Tr}\ T_{QTM}(u).
\end{equation}
It can be read off Figure \ref{startingpoint2} that 
the partition function can be evaluated as
\begin{equation}
  \label{barz2}
  \bar Z(N,L)=\bra{\Psi_+(\omega)}t_{QTM}^L(0)\ket{\Psi_-(\omega)},
\end{equation}
where $\bra{\Psi_+(\omega)}$ and $\ket{\Psi_-(\omega)}$ are initial
and final states given by the boundary conditions. 
They are tensor product of two site blocks:
\begin{equation}
  \label{vectors}
  \ket{\Psi_{-}(\omega)}=\otimes_{j=1}^N \ket{\psi_{-}(\omega)}  \qquad
   \bra{\Psi_{+}(\omega)}=\otimes_{j=1}^N \bra{\psi_{+}(\omega)},
\end{equation}
where the local two-site states are defined as
\begin{equation}
  \label{twosite}
  \begin{split}
  \ket{\psi_-(\omega)}&=\sum_{j_1,j_2=1}^2
  \psi_-(\omega,j_1,j_2)\ \ket{j_2}_2\otimes\ket{j_1}_1\\
    \bra{\psi_+(\omega)}&=\sum_{j_1,j_2=1}^2
  \psi_+(\omega,j_1,j_2)\ \bra{j_2}_2\otimes\bra{j_1}_1.
  \end{split}
\end{equation}
The components can be read off \eqref{kist}-\eqref{Tcross}:
\begin{equation}
  \label{vectors2}
  \psi_-(\omega,j_1,j_2)=(\sigma^yK_-(-\omega))_{j_1}^{j_2}\qquad
  \psi_+(\omega,j_1,j_2)=(K_+(-\omega)\sigma^y)_{j_2}^{j_1}.
\end{equation}
Here the convention is such that the index $j_1$ refers to the lower
lines in Fig  \ref{startingpoint2}.

It is important that the two boundary
states are not identical to each other, because the $K$-matrices are
evaluated at different rapidity parameters. The crossing relation
\eqref{Bcrossing} relates the two boundary states to each other, however,
this will not be used explicitly.
We also remark that the boundary
states are not normalized to unity. In fact we have
\begin{equation}
  \label{Bnorm}
  \skalarszorzat{\Psi_+(\omega)}{\Psi_-(\omega)}=
\left[  \text{Tr} \Big(K_-(-\omega)K_+(-\omega)\Big)\right]^N.
\end{equation}

Let $\langle\tilde\Psi_j|$ and $\ket{\Psi_j}$ denote the left- and
right-eigenvectors of $T_{QTM}(u)$, with the normalization
\begin{equation}
  \skalarszorzat{\tilde\Psi_j}{\Psi_k}=\delta_{jk}.
\end{equation}
The quantum transfer matrix is not
hermitian, therefore the two sets of vectors are not hermitian
conjugates of each other. They can be related to each
other by a crossing transformation, but this will not be used here.

Using the complete set of states the partition function \eqref{barz2} is evaluated as
\begin{equation}
  \bar Z(N,L)=\sum_j \Lambda_j^L(0)
  \skalarszorzat{\Psi_+(\omega)}{\Psi_j} \skalarszorzat{\tilde\Psi_j}{\Psi_-(\omega)},
\end{equation}
where $\Lambda_j(0)$ are the eigenvalues of the QTM for the special
rapidity $u=0$.
It is known that there is a finite gap in the set of the eigenvalues
even in the $N\to\infty$ limit \cite{kluemper-review}. Therefore, in
the large volume limit it is enough to keep the leading eigenvector
with the maximal eigenvalue. This state will be denoted by
$\ket{\Psi_0}$, with the eigenvalue being $\Lambda_0(0)$.

Using the relation \eqref{Z1} we get
\begin{equation}
  \begin{split}
  e^{-F/T}=&
e^{-\frac{1}{\Delta T}+\frac{2\Delta L}{T}}
\times\\
&\times \lim_{N\to\infty} \left[
\frac{2^N \sinh^{-2LN}(\eta)  \Lambda_0^L
\skalarszorzat{\Psi_{+}(\omega)}{\Psi_0}
 \skalarszorzat{\tilde\Psi_0}{\Psi_{-}(\omega)}}
    {(\text{Tr}(K^+(0))\text{Tr}(K^-(0)))^N}
  \right]+\ordo(e^{-\kappa L}),
  \end{split}
\end{equation}
where $\kappa$  is the gap in the eigenvalues in the Trotter limit.

Isolating the $\ordo(L)$ and $\ordo(1)$ terms we get
\begin{equation}
  \label{ordoL}
  e^{-f/T}= e^{2\Delta/T}  \lim_{N\to\infty}  \frac{\Lambda_0}{\sinh^{2N}(\eta)},
\end{equation}
and for the boundary free energy
\begin{equation}
  \label{FB1}
  \begin{split}
 e^{- 2F_B/T}=&
e^{-\frac{1}{\Delta T}}
 \lim_{N\to\infty} \left[
\left(\frac{2}{\text{Tr}(K^+(0))\text{Tr}(K^-(0))}\right)^N
  \skalarszorzat{\Psi_{+}(\omega)}{\Psi_0}
\skalarszorzat{\tilde\Psi_0}{\Psi_{-}(\omega)}
  \right].
  \end{split}
\end{equation}

\subsection{Bethe Ansatz for the QTM}

In order to compute \eqref{FB1} we need a description of the dominant
eigenvector of the QTM. This can be given by the Algebraic Bethe
Ansatz (ABA),
which will be summarized below.
 For a more detailed exposition we refer the reader to
\cite{QTM1,kluemper-review}. It is important that the diagonalization
of the QTM is completely independent of the boundary conditions of the
physical spin chain, and it is always done by the standard techniques
of ABA that apply to a periodic (inhomogeneous) spin chain.

We write the monodromy matrix in auxiliary space as the block-matrix
\begin{equation}
  T_{QTM}(u)=
  \begin{pmatrix}
    A(u) & B(u) \\ C(u) & D(u)
  \end{pmatrix}.
\end{equation}
Right eigenstates of the transfer matrix can be constructed as
\begin{equation}
  \label{psiB}
  \ket{\Psi}\sim \prod_{k=1}^n B(\lambda_k) \ket{0}.
\end{equation}
Here $\ket{0}$ is the reference state with all spins up, and the
$\lambda_k$ are the Bethe rapidities. The reference state is
annihilated by the $C$-operators and it is an eigenvector of $A$ and
$D$ with the eigenvalues
\begin{equation*}
  a(u)=(\sinh(u+\omega)\sinh(u-\omega+\eta))^N\qquad\qquad
  d(u)=(\sinh(u-\omega)\sinh(u+\omega-\eta))^N.
\end{equation*}
A state \eqref{psiB} is an eigenstate if
it satisfies the Bethe equations
\begin{equation}
  \label{QTMBethe}
 \fa(\lambda_j)+1=0, \qquad j=1\dots n,
\end{equation}
where we defined the auxiliary function as
\begin{equation}
\label{adef}
\begin{split}
\fa(\lambda)=&\frac{d(\lambda)}{a(\lambda)}\prod_{k=1}^n 
\frac{\sinh(\lambda-\lambda_k+\eta)}{\sinh(\lambda-\lambda_k-\eta)}=\\
=&\left(\frac{\sinh(\lambda-\eta+\omega) }{\sinh(\lambda+\eta-\omega)}
\frac{\sinh(\lambda-\omega)}{\sinh(\lambda+\omega)}\right)^N
\prod_{k=1}^n
\frac{\sinh(\lambda-\lambda_k+\eta)}{\sinh(\lambda-\lambda_k-\eta)}.
\end{split}
\end{equation}
As we will see below, the function $\fa(u)$ has a finite Trotter
limit for the ground state of the QTM. In the formulas below we will use the same notation $\fa(u)$
for finite $N$ and also for $N\to\infty$.

The corresponding eigenvalue of the transfer matrix is
\begin{equation}
  \label{TQTMu}
   \Lambda(u)=a(u)\prod_{j=1}^n
  \frac{\sinh(\lambda_j-u+\eta)}{\sinh(\lambda_j-u)}+
d(u)
\prod_{j=1}^n
  \frac{\sinh(\lambda_j-u-\eta)}{\sinh(\lambda_j-u)}.
\end{equation}
At the special rapidity $u=0$ this gives
\begin{equation}
\label{TQTM0}
\Lambda(0)=
(\sinh(\omega)\sinh(\eta-\omega))^N\left(
\prod_{j=1}^n
  \frac{\sinh(\lambda_j+\eta)}{\sinh(\lambda_j)}+
\prod_{j=1}^n
  \frac{\sinh(\lambda_j-\eta)}{\sinh(\lambda_j)}\right).
\end{equation}
Similar to \eqref{psiB}, left-eigenvectors can be constructed as
\begin{equation}
  \label{psiC}
  \bra{\Psi}\sim \bra{0}\prod_{k=1}^n C(\lambda_k).
\end{equation}
The normalization factors associated to these vectors are  \cite{korepin-norms}
\begin{equation}
  \label{norm}
  \bra{0}\prod_{k=1}^n C(\lambda_k)  \prod_{k=1}^n B(\lambda_k) \ket{0}=
\sinh^n(\eta) \prod_{j=1}^n (a(\lambda_j)d(\lambda_j))
\prod_{j\ne k} f(\lambda_j,\lambda_k) \times \det G^{[n]},
\end{equation}
where
\begin{equation}
  f(\lambda,\mu)=\frac{\sinh(\lambda-\mu+\eta)}{\sinh(\lambda-\mu)}
\end{equation}
and $G^{[n]}$ is the so-called Gaudin matrix given by
\begin{equation}
  \label{Gaudin}
  G_{jk}^{[n]}=\delta_{jk}m_j-\varphi(\lambda_j-\lambda_k)
   \end{equation}
 with 
 \begin{equation}
   \label{mdef}
  m_j=\frac{\fa'(\lambda_j)}{\fa(\lambda_j)}
\end{equation}
and
\begin{equation}
  \varphi(\lambda) =- \frac{\sinh(2\eta)}{\sinh(\lambda+\eta)\sinh(\lambda-\eta)}.
\end{equation}

It is known that the leading eigenvector of the QTM has $n=N$ rapidities which
come in pairs and can be denoted as
\begin{equation}
  \{\lambda\}_N=\{\lambda^+,-\lambda^+\}_{N/2}=
\{\lambda_1^+,-\lambda_1^+,\lambda_2^+,-\lambda_2^+,\dots,\lambda_{N/2}^+,-\lambda_{N/2}^+\}.
\end{equation}
This holds for arbitrary temperatures. In the massive regime all $\lambda^+$ are purely imaginary, whereas
for $\Delta<1$ they are real. For a more detailed discussion of the
Bethe roots of the QTM we refer the reader to \cite{kluemper-review}.

The advantage of the auxiliary function is that sums over the Bethe
roots can be expressed as contour integrals involving
$\fa(u)$, and that $\fa(u)$ has a finite $N\to\infty$ limit. Moreover,
there are simple non-linear integral equations (NLIE's) that determine
$\fa(u)$ both at finite $N$ and in the Trotter limit. 
In the remainder of this section we repeat the main steps in deriving
these NLIE's. Essentially the same steps will be used
later to derive the integral representations for the boundary free energy.

First we note that both $\fa(u)$ and $\Lambda(u)$ are periodic in the
imaginary direction with period $i\pi$. Second, we define the
so-called physical strip. In the massive case this consists of the
points of complex plane with $|\Re(z)|<\eta/2$, $|\Im(z)|<\pi/2$,
whereas in the massless case we require $|\Im(z)|<\gamma/2$. The most
important property of the physical strip is that the only zeroes of
$1+\fa(u)$ within the strip are the Bethe roots, and this is why
we can use the contour integral techniques for the sums over Bethe roots \cite{kluemper-review}.

As customary we define a contour $\CC$ such that it encircles all
Bethe roots and the $N$-order pole of $\fa(\lambda)$ at
$\lambda=-\omega$, but it does not include additional poles or zeroes
of $1+\fa(\lambda)$ \cite{kluemper-review}.  As remarked above, this
means that $\CC$ has to stay in the physical strip.
In the massive regime the contour can be chosen as the union of two
vertical segments $[-i\pi/2+a\dots i\pi/2+a]$ and
$[-i\pi/2-a\dots i\pi/2-a]$ with $a<\eta/2$. In the massless regime
the usual choice is the union of $\valos+ia$ and $\valos-ia$ with $a<\gamma/2$, but for
our present purposes it is important to choose a compact contour. One
possibility is to choose an ellipse whose axes coincide with the real
and imaginary axis, such that the minor axis is smaller than
$\gamma/2$ and the major axis is large enough to include all
Bethe roots; this is our choice for the numerical investigations, to
be presented in Sec. \ref{sec:numerics}.

As a first step we express the logarithm of $\fa(\lambda)$ as
\begin{equation}
\begin{split}
\log \fa(\lambda)=&N\log \left(\frac{\sinh(\lambda+\eta+\omega) }{\sinh(\lambda+\eta-\omega)}
\frac{\sinh(\lambda-\omega)}{\sinh(\lambda+\omega)}\right)-
N\log
\frac{\sinh(\lambda+\eta+\omega)}{\sinh(\lambda-\eta+\omega)}+\\
&+\sum_{k=1}^N\log \frac{\sinh(\lambda-\lambda_k+\eta)}{\sinh(\lambda-\lambda_k-\eta)}.
\end{split}
\end{equation}
Within the contour $\mc{C}$ the function $1+\fa(u)$ has an $N$-fold
pole at $u=-\omega$ and $N$ zeroes at the Bethe roots. Therefore
$\log(1+\fa(u))$ can be defined as a single valued function on the
contour. 
Let us choose a function $f$ that does not have poles or zeros within the
contour. Then from integration by parts we get the formula
\begin{equation}
  \log\pa{\pl{a=1}{N}f\pa{\la_a}}=
  -\oint\limits_{\mc{C}}\f{\mathrm{d} \nu}{2\pi i}\f{f'\pa{\nu}}{f\pa{\nu}}\log\pa{1+\mf{a}(\nu)}
+N\log f\pa{-\omega}.
\label{integral}
\end{equation}
Applying this identity with
$f(\nu)=\frac{\sinh(\nu-\lambda+\eta)}{\sinh(\nu-\lambda-\eta)}$
we get
\begin{equation}
\begin{split}
\log \fa(\lambda)=&N\log \left(\frac{\sinh(\lambda+\eta+\omega) }{\sinh(\lambda+\eta-\omega)}
\frac{\sinh(\lambda-\omega)}{\sinh(\lambda+\omega)}\right)
+  \int_C \frac{d\nu}{2\pi i} \varphi(\lambda-\nu)\log(1+\fa(\nu)).
\end{split}
\end{equation}
This integral equation is valid at any finite Trotter number $N$ and for any parameter
$\omega$. 
Also, it has a finite Trotter limit given by
\begin{equation}
\label{NLIE}
\begin{split}
\log \fa(\lambda)=&
-\frac{1}{T}\frac{2\sinh^2(\eta)}{\sinh(\lambda)\sinh(\lambda+\eta)}
+  \int_\CC \frac{d\nu}{2\pi i} 
\varphi(\lambda-\nu) \log(1+\fa(\nu)).
\end{split}
\end{equation}

The transfer matrix eigenvalue is determined similarly. From
\eqref{TQTMu} we have
\begin{equation}
  \log  \Lambda(u)=\log(1+\fa(u))+N\log(
(\sinh(u+\omega)\sinh(u-\omega+\eta))
  )+  \sum_{j=1}^N
\log  \frac{\sinh(\lambda_j-u+\eta)}{\sinh(\lambda_j-u)}.
\end{equation}
Applying \eqref{integral} with
$f(\nu)=\frac{\sinh(\nu-u+\eta)}{\sinh(\nu-u)}$ with $u$ outside the
contour we get
\begin{equation}
  \begin{split}
  \log  \Lambda(u)=&\log(1+\fa(u))+N\log(
(\sinh(\omega+u-\eta)\sinh(u-\omega+\eta))
)+\\
&+\oint\limits_{\mc{C}}\f{\mathrm{d} \nu}{2\pi  i}
\frac{\sinh(\eta)}{\sinh(\nu-u+\eta)\sinh(\nu-u)}
\log\pa{1+\mf{a}(\nu)}.
\end{split}
\end{equation}
We can analytically continue this formula to the case when $u$ is
inside the contour. 
In this case
we pick up a pole
contribution and (using also that $N$  is even)
\begin{equation}
  \begin{split}
  \log  \Lambda(u)=&N\log(
(\sinh(-\omega-u+\eta)\sinh(u-\omega+\eta))
)+\\
&+\oint\limits_{\mc{C}}\f{\mathrm{d} \nu}{2\pi  i}
\frac{\sinh(\eta)}{\sinh(\nu-u+\eta)\sinh(\nu-u)}
\log\pa{1+\mf{a}(\nu)}.
\end{split}
\end{equation}
Specifically for $u=0$
\begin{equation}
  \begin{split}
  \log  \Lambda(0)=&2N\log(
\sinh(-\omega+\eta))
)+\oint\limits_{\mc{C}}\f{\mathrm{d} \nu}{2\pi  i}
\frac{\sinh(\eta)}{\sinh(\nu+\eta)\sinh(\nu)}
\log\pa{1+\mf{a}(\nu)}.
\end{split}
\end{equation}
Thus for the free energy density we get from \eqref{ordoL}
\begin{equation}
  \label{fNLIE}
  \begin{split}
f/T&= -2\Delta/T-\lim_{N\to\infty} 2N\log\frac{\sinh(-\omega+\eta)}{\sinh(\eta)}
-\oint\limits_{\mc{C}}\f{\mathrm{d} \nu}{2\pi  i}
\frac{\sinh(\eta)}{\sinh(\nu+\eta)\sinh(\nu)}
\log\pa{1+\mf{a}(\nu)}=\\
&
=-\oint\limits_{\mc{C}}\f{\mathrm{d} \nu}{2\pi  i}
\frac{\sinh(\eta)}{\sinh(\nu+\eta)\sinh(\nu)}
\log\pa{1+\mf{a}(\nu)}.
\end{split}
\end{equation}

In the calculation of the boundary free energy the special rapidity
 $i\pi/2$ will be of importance. From the definition \eqref{adef}
it can be seen that $\fa(i\pi/2)=1$ for any $\omega$ and $N$. This
value is thus constant even in the Trotter limit, for any 
temperature.  

\section{Scalar products}

\label{sec:overlaps}

In this section we treat the 
the scalar products between the eigenstates of the QTM and the
boundary states. Our goal is to find convenient determinant formulas
for the products
\begin{equation}
   \skalarszorzat{\Psi_{+}(\omega)}{\Psi}
\skalarszorzat{\tilde\Psi}{\Psi_{-}(\omega)},
\end{equation}
which enter formula \eqref{FB1}. We stress that the Bethe states are
normalized to unity, but the boundary states have a non-trivial norm
given by \eqref{Bnorm}. According to the previous section, such
products can be evaluated in the Algebraic Bethe Ansatz as
\begin{equation}
  \frac{\bra{\Psi_{+}(\omega)} \prod_{k=1}^n B(\lambda_k) \ket{0}
    \bra{0}\prod_{k=1}^n C(\lambda_k) \ket{\Psi_{-}(\omega)}}
{ \bra{0}\prod_{k=1}^n C(\lambda_k)  \prod_{k=1}^n B(\lambda_k) \ket{0}}.
\end{equation}
The un-normalized overlaps in the numerator are related to each other as
\begin{equation}
  \label{ovcross}
\bra{\Psi_{+}(\omega)} \prod_{k=1}^n B(\lambda_k) \ket{0}=
\left(\frac{\sinh(2(\eta-\omega))}{\sinh(-2\omega)}\right)^n    \bra{0}\prod_{k=1}^n C(\lambda_k) \ket{\Psi_{-}(\omega)}.
\end{equation}
This relation was already given in \cite{sajat-karol}, and it can be
proven using the crossing relations
\eqref{Rcross} and \eqref{Bcrossing}.

Scalar products of this form were first studied in \cite{sajat-neel},
where it was shown that they are related to a certain partition
function of the six-vertex model with a reflecting end. In the case
of diagonal $K$-matrices the partition function was computed by Tsushiya 
in \cite{tsushiya}, the
resulting determinant formula is known as the Tsushiya-determinant. 
An alternative determinant formula for the overlaps was given in
\cite{sajat-karol}. This was used as a starting point in the works 
\cite{Caux-Neel-overlap1,Caux-Neel-overlap2} that considered the
overlaps in the homogeneous, physical spin chain. In these works it was shown that
the only non-vanishing overlaps are those with states that display the
pair structure, and that in these cases the scalar products can be
expressed by two Gaudin-like determinants multiplied by a product of
one-particle overlap functions. The formulas of
\cite{Caux-Neel-overlap1,Caux-Neel-overlap2} proved to be central for
the analysis of quantum quench problems in the spin chain
\cite{JS-oTBA,sajat-oTBA,JS-CGGE}.

The recent work \cite{sajat-minden-overlaps} extended the
results of \cite{Caux-Neel-overlap1,Caux-Neel-overlap2} to boundary
states with non-diagonal K-matrices. It was found that the overlaps
have the same structure even in this case: they involve the same
Gaudin-like determinants, and only the pre-factors
need to be modified. It was argued in \cite{sajat-minden-overlaps} that the
pre-factors can be fixed by a combination of the QTM and
Quench Action methods, but the presence of the same determinants was
only conjectured and numerically checked; a proof is yet missing. 

In \cite{sajat-integrable-quenches} it was shown by a very general, 
model-independent calculation that the overlaps with
boundary state created by integrable $K$-matrices are non-vanishing
only for Bethe states with the pair structure. This proof did not rely on knowledge of the exact
overlaps, instead it only used the boundary Yang-Baxter relation
\eqref{BYB}. Therefore it also applies to the non-diagonal $K$-matrices.

The works \cite{sajat-neel,Caux-Neel-overlap1,Caux-Neel-overlap2,sajat-integrable-quenches,sajat-minden-overlaps} studied the homogeneous chain which corresponds to 
 $\omega=\eta/2$ in \eqref{TQTM}. Fewer results are available for the
 inhomogeneous case, which can  nevertheless be studied with the same
 methods. For example it can be seen easily that the proof of
 \cite{sajat-integrable-quenches} for the pair structure holds for
 general $\omega$. Moreover, the methods of
 \cite{sajat-minden-overlaps} can be modified in a straightforward way
 to yield a conjecture for the overlaps relevant to this work. 
 In the following we formulate this conjecture, whereas for a detailed
 calculation leading up to it we refer to \cite{sajat-minden-overlaps}.

Denoting by $\ket{\{\pm\lambda^+\}_{n/2}}$  the on-shell states with the pair structure
the conjecture for the overlaps reads
\begin{equation}
  \begin{split}
  &  \skalarszorzat{\Psi_{+}(\omega)}{\{\pm\lambda^+\}_{n/2}}
\skalarszorzat{\{\pm\lambda^+\}_{n/2}}{\Psi_{-}(\omega)}
=  \\ &
 \hspace{3cm}=(\sinh(2(\eta-\omega))\sinh(2\omega))^{N}
\prod_{j=1}^{n/2}  u(\lambda^+_j)
\times \frac{\det G_{jk}^{+,[n/2]}}{\det G_{jk}^{-,[n/2]}}.
\end{split}
  \label{OVERLAPS3}
\end{equation}
Here the single particle overlap function is defined as
\begin{equation}
  \label{ulambda2}
 u(\lambda)=\frac{  (v_{\alpha}^sv_{\beta}^c)^2}
  {v^s_{\eta/2} v^c_{\eta/2}v^s_0v^c_0},
\end{equation}
where we used the
short-hand notation
\begin{equation}
  v^s_\kappa(\lambda)=\sinh(\lambda+\kappa)\sinh(\lambda-\kappa)\qquad
   v^c_\kappa(\lambda)=\cosh(\lambda+\kappa)\cosh(\lambda-\kappa).
 \end{equation}
The two Gaudin-like matrices in \eqref{OVERLAPS3} are
\begin{equation}
  G^{\pm,[n/2]}_{jk}=\delta_{jk}m_j
  -\left[\varphi(\lambda^+_j-\lambda^+_k)\pm \varphi(\lambda^+_j+\lambda^+_k)\right].
 \end{equation}

 The differences between the present formula \eqref{OVERLAPS3} and
the main result in \cite{sajat-minden-overlaps}  (eq. (3.13) there)  are the
following:
\begin{itemize}
  \item In the present case the length of the (auxiliary) spin chain 
    is $2N$, because is the overlap is calculated in the channel of
    the Quantum Transfer Matrix. In \cite{sajat-minden-overlaps} the
    length of the spin chain was denoted by $L$, and it referred to the
    physical spin chain. 
  \item Here the boundary states are not normalized;
    overlaps with normalized states are obtained by dividing with the
factor given by \eqref{Bnorm}.
\item The Gaudin-like matrices $G^\pm$ are expressed in terms of the
  auxiliary function: the diagonal elements $m_j$ are given by
  \eqref{mdef}. This functional form holds for any $\omega$, 
  including the homogeneous case. 
 However, $\fa(\lambda)$ depends
  explicitly on $\omega$, thus the diagonal elements of  $G^\pm$
  differ from those in  \cite{Caux-Neel-overlap1,Caux-Neel-overlap2,sajat-minden-overlaps}.
\item Here we have a non-trivial overall normalization factor of
  $(\sinh(2(\eta-\omega))\sinh(2\omega))^{N}$.  This factor is found
  by evaluating the ``Quench Action sum rule'', analogously to the
  calculation in  \cite{sajat-minden-overlaps}. The overlap formula
  also holds for $n=0$, in which case this factor
  comes simply from the projection of the reference state onto the
  boundary states. These overlaps are obtained simply from
  \eqref{vectors}-\eqref{twosite} after the substitution
  \begin{equation}
    \begin{split}
    \psi_-(-\omega)_{11}&=-i K_{-}(-\omega)_{12}=i\sinh(2\omega)\\
      \psi_+(-\omega)_{11}&=i
      K_{+}(-\omega)_{21}=i\sinh(2(\eta-\omega)).
      \end{split}
  \end{equation}
\end{itemize}

These differences can be summarized as follows: After the replacements $2N\to L$ and $\omega\to \eta/2$ (and
correcting with the norm \eqref{Bnorm}) the formula
\eqref{OVERLAPS3} reproduces the main result of
\cite{sajat-minden-overlaps}. 

The case of diagonal $K$-matrices can be obtained in the limit
$\beta\to\infty$ by the scaling
\begin{equation*}
  K_{\text{diag}}(u)=\lim_{\beta\to\infty} e^{-\beta} K(u)=
  \begin{pmatrix}
    \sinh(\alpha+u) & 0 \\ 0 & \sinh(\alpha-u)
  \end{pmatrix}.
\end{equation*}
It can be seen from \eqref{OVERLAPS3} that in this limit the un-normalized overlaps scale as
$e^{2\beta n}$, therefore, only the Bethe states with $n=N$ have a
non-vanishing overlap with the boundary states created by the diagonal
$K$-matrices \footnote{Here we used the fact that the
  Algebraic Bethe Ansatz only treats states with $n\le N$.}. As
remarked in \cite{sajat-minden-overlaps}, this is consistent with
spin-$z$ conservation, because according to \eqref{twosite} the diagonal $K$-matrices describe
states that lie in the zero-magnetization sector.

In the diagonal case our conjecture 
can be proven
rigorously using the methods of
\cite{Caux-Neel-overlap1,Caux-Neel-overlap2}, starting from the
Tsushiya-determinant and its alternative representation given in
\cite{sajat-karol}. 
In
\cite{Caux-Neel-overlap1,Caux-Neel-overlap2} only the homogeneous case
was treated, in order to arrive at the overlaps with the N\'eel state
for the physical spin chain. However, each step can be repeated in a
straightforward way also for the inhomogeneous transfer matrices,
incorporating the relations \eqref{ovcross} and \eqref{norm}.
We performed this
calculation and thus proved
\eqref{OVERLAPS3} in the diagonal case; we refrain from presenting
this calculation here, because it is a mere repetition
of the steps of \cite{Caux-Neel-overlap1,Caux-Neel-overlap2}.

We also performed numerical
checks of our conjecture in the case of general boundary parameters;
we treated cases
with small values of $N$ and $\Delta>1$. The numerical procedure included the solution of
the Bethe equations \eqref{QTMBethe}, which is relatively easy as all
roots lie on a finite segment on the imaginary axis. The numerical
predictions of \eqref{OVERLAPS3} were then compared to real space
calculation of the overlaps, for various values of the boundary
parameters. In all cases perfect agreement was found,
with a relative accuracy of at least $10^{-10}$. Examples of the
numerical data are shown in Table \ref{tab:ov}.

\begin{table}
  \centering
  \begin{tabular}{|c|l|c|}
    \hline
   $N$ &$\{\text{Im}(\lambda^+)\}_{N/2}$&  $S$\\
    \hline
    \hline
    2 & $\{$0.337728$\}$ & 63.022646 \\
    \hline
    4 & $\{$0.146496,0.620620$\}$ &  2773.2102 \\
       \hline
    6 & $\{$0.0711016,0.2514235,0.6705907$\}$ &  30066.783 \\
    \hline   
  \end{tabular}
  \caption{Numerical examples for the overlaps \eqref{OVERLAPS3} for
    the leading eigenstate of the QTM, for
    small values of $N$. We denote $S=\skalarszorzat{\Psi_{+}(\omega)}{\{\pm\lambda^+\}_{N/2}}
\skalarszorzat{\{\pm\lambda^+\}_{N/2}}{\Psi_{-}(\omega)}$, where the
left and right eigenstates are normalized to
$\skalarszorzat{\{\pm\lambda^+\}_{N/2}}{\{\pm\lambda^+\}_{N/2}}=1$ and
the boundary states are given by \eqref{vectors}. In these examples we
chose $\Delta=2$ such that the rapidities are purely imaginary. We
chose the arbitrary parameters $T=1$, $\alpha=0.1$ and
$\beta=1.5$. The relative error of the formula \eqref{OVERLAPS3}  was
smaller than $10^{-10}$ in all three cases. Similar errors were
observed for other values of $\Delta$ and the boundary parameters.}
  \label{tab:ov}
\end{table}

\section{Boundary free energy}

\label{sec:final}

In this Section we compute the Trotter limit of the previous formulas
and derive a compact representation for the boundary free energy. We
treat the leading eigenstate of the QTM, whose rapidities will be
denoted as
\begin{equation*}
  \{\lambda\}_N=\{\lambda^+,-\lambda^+\}_{N/2}=\{\lambda_1^+,-\lambda_1^+,\lambda_2^+,-\lambda_2^+,\dots,\lambda_{N/2}^+,-\lambda_{N/2}^+\}.
\end{equation*}
Starting from \eqref{FB1} and \eqref{OVERLAPS3}, and 
computing the traces of the $K$-matrices as
\begin{equation}
  \text{Tr}(K^-(0))=4\sinh(\alpha)\cosh(\beta)\qquad
   \text{Tr}(K^+(0))=4\sinh(\alpha)\cosh(\beta)\cosh(\eta)
\end{equation}
we obtain
\begin{equation}
  \begin{split}
&  e^{-2\beta  F_B}=e^{-\frac{1}{\cosh(\eta)T}}\lim_{N\to\infty}   \left(\frac{\cosh(\eta-\omega)}{\cosh(\eta)}\right)^N
\left(  \frac{\sinh(\eta-\omega)}{\sinh(\eta-2\omega)}\right)^N\times\\
&  \prod_{j=1}^N\left[ \frac{\sinh^2(\lambda_j+\alpha)\cosh^2(\lambda_j+\beta)}
  {\sinh^2(\alpha)\cosh^2(\beta)}
  \frac{\sinh(2\omega)}{\sinh(2\lambda_j)}
  \frac{\sinh(\eta-2\omega)}{\sinh(2\lambda_j+\eta)}\right]\times
\frac{\det  G^{+,[N/2]}}{\det G^{-,[N/2]}}.
\end{split}
\end{equation}
The Trotter limit of the first two pre-factors is 
\begin{equation}
  \begin{split}
&\lim_{N\to\infty}   \left(\frac{\cosh(\eta-\omega)}{\cosh(\eta)}\right)^N
\left(  \frac{\sinh(\eta-\omega)}{\sinh(\eta-2\omega)}\right)^N
=e^{\frac{1}{\cosh(\eta)}\frac{1}{T}}
\end{split}
\end{equation}
leading to the simplified expression
\begin{equation}
  \label{interm}
  \begin{split}
    &  e^{-2\beta  F_B}=
 \lim_{N\to\infty}  \prod_{j=1}^N\left[ \frac{\sinh^2(\lambda_j+\alpha)\cosh^2(\lambda_j+\beta)}
  {\sinh^2(\alpha)\cosh^2(\beta)}
  \frac{\sinh(2\omega)}{\sinh(2\lambda_j)}
  \frac{\sinh(\eta-2\omega)}{\sinh(2\lambda_j+\eta)}\right]\times
\frac{\det  G^{+,[N/2]}}{\det G^{-,[N/2]}}.
\end{split}
\end{equation}

The Gaudin-like matrices are convenient for a numerical analysis at
finite Trotter number, but they are somewhat difficult to treat
in the Trotter limit. Our goal is to express them as Fredholm
determinants, therefore we rewrite them as $N\times N$ matrices.
Let us define the $N\times N$ matrices
$H^\pm_{jk}$ as 
\begin{equation}
  H^{\pm,[N]}_{jk}= \frac{\varphi_{jk}^{-,[N]}\pm \varphi_{jk}^{+,[N]}}{2},
\text{ where }\quad  \varphi_{jk}^{\pm,[N]}=\frac{\varphi_\eta(\lambda_j\pm \lambda_k)}{m_j}.
\end{equation}
It can be seen from the block diagonal forms that
\begin{equation*}
  \frac{\det G^{+,[N/2]}}{\det G^{-,[N/2]}}=
   \frac{\det\left(1- H^{+,[N]}\right)}{\det\left(1- H^{-,[N]}\right)},
\end{equation*}
and this expression is more convenient in the subsequent calculations.

\subsection{Trotter limit}

We are now in the position to perform the Trotter limit. First we
consider the pre-factors: we apply \eqref{integral} to turn the products into integral
expressions.

Let us consider the function $f(u)=\sinh(u+\alpha)$ with a
boundary parameter $\alpha$ that lies outside the canonical contour. Then from 
\eqref{integral} we get
\begin{equation}
  \log \pa{\pl{a=1}{N}\frac{\sinh \pa{\la_a+\alpha}}{\sinh(\alpha)}}=
  -\oint\limits_{\mc{C}}\f{\mathrm{d} \nu}{2\pi i}
\coth(\nu+\alpha)
  \log\pa{1+\mf{a}(\nu)}
+N\log \frac{\sinh\pa{-\omega+\alpha}}{\sinh(\alpha)}.
\label{integralxi}
\end{equation}
In the case when $\alpha$ is inside the contour we have to pick up a pole
contribution, leading to the general formula
\begin{equation}
  \begin{split}
&  \log \pa{\pl{a=1}{N}\frac{\sinh \pa{\la_a+\alpha}}{\sinh(\alpha)}}=
  \delta_{-\alpha}\log(1+\fa(-\alpha))-\\
&\hspace{2cm}  -\oint\limits_{\mc{C}}\f{\mathrm{d} \nu}{2\pi i}
\coth(\nu+\alpha)
  \log\pa{1+\mf{a}(\nu)}
+N\log \frac{\sinh\pa{-\omega+\alpha}}{\sinh(\alpha)},
\label{integralxii}
\end{split}
\end{equation}
where we used the notation that for any $\xi\in\complex$ we have $\delta_{\xi}=1$ if $\xi$ is inside
the contour, and $\delta_{\xi}=0$ otherwise.

In the Trotter limit this leads to
\begin{equation}
  \begin{split}
&\lim_{N\to\infty}  \log \pa{\pl{a=1}{N}\frac{\sinh \pa{\la_a+\alpha}}{\sinh(\alpha)}}=
  \delta_{-\alpha}\log(1+\fa(-\alpha))-\\
&\hspace{2cm}  -\oint\limits_{\mc{C}}\f{\mathrm{d} \nu}{2\pi i}
\coth(\nu+\alpha)
\log\pa{1+\mf{a}(\nu)}-
\frac{\coth(\alpha)\sinh(\eta)}{T}.
\label{integralxalpha}
\end{split}
\end{equation}
Similarly
\begin{equation}
  \begin{split}
&\lim_{N\to\infty}  \log \pa{\pl{a=1}{N}\frac{\cosh \pa{\la_a+\beta}}{\cosh(\beta)}}=
  \delta_{-\beta-i\pi/2}\log(1+\fa(-\beta-i\pi/2))-\\
&\hspace{2cm}  -\oint\limits_{\mc{C}}\f{\mathrm{d} \nu}{2\pi i}
\tanh(\nu+\beta)
\log\pa{1+\mf{a}(\nu)}-
\frac{\tanh(\beta)\sinh(\eta)}{T}.
\label{integralxbeta}
\end{split}
\end{equation}

By analogous calculations for the remaining two products we obtain
\begin{equation}
  \begin{split}
&    \lim_{N\to\infty}  \log \prod_{a=1}^N   \frac{\sinh(2\omega)}{\sinh(2\lambda_j)}
    \frac{\sinh(\eta-2\omega)}{\sinh(2\lambda_j+\eta)}=\\
  & \hspace{2cm} =-\log(2)(1+ \delta_{i\pi/2})+
  \oint\limits_{\mc{C}}\f{\mathrm{d} \nu}{2\pi i}
2(\coth(2\nu)+\coth(2\nu+\xi))
  \log\pa{1+\mf{a}(\nu)}.
  \end{split}
\end{equation}
Here we used that at finite Trotter number we have the special values $\fa(0)=\fa(i\pi/2)=1$.
In the massive case $\delta_{i\pi/2}=1$, whereas in the massless case
$\delta_{i\pi/2}=0$, therefore we will use the notation
$\delta_{\Delta>1}\equiv \delta_{i\pi/2}$.

We now evaluate the Trotter limit of the determinants.  For any matrix of the form
\begin{equation}
M^{[N]}_{jk}=  \delta_{jk}-\frac{\fa(\lambda_j)}{\fa'(\lambda_j)}f(\lambda_j,\lambda_k)
\end{equation}
its determinant in the Trotter limit becomes the Fredholm determinant
\begin{equation}
  \label{Fred}
\lim_{N\to\infty} M^{[N]}=\det\left(1-\hat f\right),
\end{equation}
where $\hat f$ is an integral operator that acts as
\begin{equation}
  \left(\hat f (g)\right)(x)=\int_{\mathcal{C}} \frac{du}{2\pi i}
  \frac{\fa(u)}{1+\fa(u)} f(x,u) g(u)
\end{equation}
This can be applied to the matrices above, so that
\begin{equation}
  \lim_{N\to\infty}   \frac{\det\left(1 - H^{+,[N]}\right) }{\det\left(1- H^{-,[N]}\right) }=
  \frac{\det\left(1-\hat H^{+}\right)}{\det\left(1-\hat H^-\right)},
\end{equation}
where $\hat H^{\pm}$
are integral operators that
act as
\begin{equation}
  \label{kernels}
  \begin{split}
    \left(\hat H^{\pm} (g)\right)(x)&=\int_{\mathcal{C}} \frac{du}{2\pi i}
  \frac{\fa(u)}{1+\fa(u)} \frac{\varphi(x-u)\pm\varphi(x+u)}{2} g(u).
\end{split}
\end{equation}
In establishing the relation \eqref{Fred} it is essential that the
only zeros of $1+\fa(u)$ inside the contour are the Bethe roots.

Putting everything together we obtain the final formula for the boundary free energy
\begin{equation}
  \label{final}
  \begin{split}
& F_B/T=
  \frac{1+\delta_{\Delta>1}}{2}\log(2)
 -\delta_{-\alpha}\log(1+\fa(-\alpha))-\delta_{-\beta-i\pi/2}\log(1+\fa(-\beta-i\pi/2))-
 \\
 & -\oint\limits_{\mc{C}}\f{\mathrm{d} \nu}{2\pi i}
 \left[\coth(2\nu)+\coth(2\nu+\eta)-\coth(\nu+\alpha)-\tanh(\nu+\beta)
   \right]\log(1+\fa(\nu))-\\
   &
   -\frac{1}{2}\log
    \frac{\det\left(1-\hat H^{+}\right)}{\det\left(1-\hat H^-\right)}
 +
\frac{(\coth(\alpha)+\tanh(\beta))\sinh(\eta)}{T}.
\end{split}
\end{equation}

An interesting alternative formula can be derived if we express the
ratio of the two determinants at finite Trotter number as
\begin{equation}
  \log   \frac{\det\left(1-\hat H^{+,[N]}\right)}{\det\left(1-\hat H^{-,[N]}\right)}=
-\sum_{n=1}^\infty \frac{1}{n} \text{Tr}\left( (H^{+,[N]})^n- (H^{-,[N]})^n\right).
\end{equation}
Using the definition of the $H^{\pm,[N]}$ and the symmetry of the Bethe roots this
is equivalent to
\begin{equation}
 - \sum_{n=1}^\infty \frac{1}{n} \text{Tr}\left(
  (\varphi^{-,[N]})^{n-1} \varphi^{+,[N]}\right),
\end{equation}
which can be expressed as the integral series
\begin{equation}
  \label{dorey-tateo}
  \begin{split}
&      \log   \frac{\det\left(1-\hat    H^{+,[N]}\right)}{\det\left(1-\hat H^{-,[N]}\right)}=\\
&-  \sum_{n=1}^\infty \frac{1}{n}
\left(\prod_{j=1}^n \int_{\CC}\frac{d\lambda_j}{2\pi i} \frac{\fa(\lambda_j)}{1+\fa(\lambda_j)}\right)
\varphi(\lambda_1-\lambda_2)\varphi(\lambda_2-\lambda_3)\dots \varphi(\lambda_{n-1}-\lambda_{n})
\varphi(\lambda_{n}+\lambda_1).
\end{split}
\end{equation}
The r.h.s. has a finite Trotter limit, which can be substituted into \eqref{final}.

Quite interestingly this expression has the same form as the
boundary-independent part of the $g$-function in the TBA framework
\cite{Dorey:2004xk,Dorey:2005ak,sajat-g,woynarovich-uj}, apart from an
overall sign. This difference in the overall sign was already noticed
in \cite{sajat-minden-overlaps}, although in a different
context (see subsection 2.5 there).

\bigskip

The physical interpretation of the terms proportional to $\delta_{-\alpha}$ and
$\delta_{-\beta-i\pi/2}$ is not evident from the finite $T$
formulas. First of all, the inclusion of these terms clearly depends
on the contour itself: even if $\alpha$ and/or $\beta$ lie within the
physical strip, the contours could be modified such that these terms
can be avoided. Their physical meaning shows up in the $T\to 0$ limit, where
they describe the contribution of  boundary bound states to the
ground state energy. Therefore we interpret them at finite $T$ as the
contribution of the boundary bound states to the free energy.
Although the $T\to 0$ limit will be investigated separately in
Section \ref{sec:T0}),
we give here a few more comments about this issue.

In the present paper we did not discuss the solution of the
Hamiltonian itself. Nevertheless it is known that the open spin chain
can have bound states localized around the boundaries, and these
states are represented by special Bethe roots. The bound states exist
when the boundary parameters are within the physical strip
\cite{skorik-saleur,skorik-kapustin}.  These bound
states modify the energy of the Bethe states through their bare energy and also by
shifting the roots of the bulk particles. Thus we can expect that they
give some finite contributions to the free energy as well.
We also expect that if a correct TBA result for $F_B$ were given, then
these terms
would show up  as the result of the existence of the 
boundary bound states.

\subsection{Special cases}

\label{ref:special}

Here we investigate special cases of the boundary parameters.

The case of longitudinal fields can be reached by sending one of the
boundary parameters to infinity. This is most easily performed on the
intermediate expression \eqref{interm}: the pair structure of
the Bethe roots implies that in the $\alpha,\beta\to\infty$ limit the
corresponding factors disappear from the product. Nevertheless it is
also possible to perform this limit on the final formulas.
For
example in the 
$\beta\to+\infty$ limit the $\beta$-dependent terms of \eqref{final}
become
\begin{equation}
  \label{betainf}
   \oint\limits_{\mc{C}}\f{\mathrm{d} \nu}{2\pi i}\log(1+\fa(\nu))
  +\frac{\sinh(\eta)}{T}.
\end{equation}
The integral can be symmetrized with a $\nu\to -\nu$ transformation,
and by using $\fa(-\nu)=1/\fa(\nu)$ we get
\begin{equation}
    \oint\limits_{\mc{C}}\f{\mathrm{d} \nu}{2\pi i}\log(1+\fa(\nu))=
    \frac{1}{2}
    \oint\limits_{\mc{C}}\f{\mathrm{d} \nu}{2\pi i}
    \log\frac{1+\fa(\nu)}{{1+\fa^{-1}(\nu)}}=
     \frac{1}{2}
    \oint\limits_{\mc{C}}\f{\mathrm{d} \nu}{2\pi i}
    \log \fa(\nu).
  \end{equation}
Using the NLIE \eqref{NLIE} this is can be evaluated by contour
integrals. The integral over the convolution term gives zero, whereas
\begin{equation}
     \frac{1}{2}
    \oint\limits_{\mc{C}}\f{\mathrm{d} \nu}{2\pi i}
    \log \fa(\nu)=  \frac{1}{2}  \oint\limits_{\mc{C}}\f{\mathrm{d} \nu}{2\pi i}
    -\frac{1}{T}\frac{2\sinh^2(\eta)}{\sinh(\lambda)\sinh(\lambda+\eta)}=-\frac{\sinh(\eta)}{T}.
\end{equation}
Altogether this shows that the sum \eqref{betainf} vanishes and for
the boundary free energy we get
\begin{equation}
  \label{diagfinal}
  \begin{split}
 F_B/T=&
  \frac{1+\delta_{\Delta>1}}{2}\log(2)
 -\delta_{-\alpha}\log(1+\fa(-\alpha))-
 \\
 & -\oint\limits_{\mc{C}}\f{\mathrm{d} \nu}{2\pi i}
 \left[\coth(2\nu)+\coth(2\nu+\eta)-\coth(\nu+\alpha)
   \right]\log(1+\fa(\nu))-\\
   &
   -\frac{1}{2}\log
 \frac{\det\left(1-\hat H^{+}\right)}{\det\left(1-\hat H^-\right)}
 +
\frac{\coth(\alpha)\sinh(\eta)}{T},
\end{split}
\end{equation}
where now $h_z=\sinh(\eta)\coth(\alpha)$.

\bigskip

It is interesting to consider the case of fixed boundary
conditions. For example, fixing the boundary spin to
$\vev{\sigma_z}=-1$ can be obtained from \eqref{diagfinal} by the $h_z\to\infty$, $\alpha\to
0$ 
limit. In this case $\alpha$ is within the
canonical contour and \eqref{diagfinal} has two diverging terms. Using
the NLIE \eqref{NLIE} we get in this limit 
\begin{equation}
  -\delta_{-\alpha}\log(1+\fa(-\alpha))+\frac{\coth(\alpha)\sinh(\eta)}{T}=
  -\frac{h_z}{T}+\ordo(\alpha).
\end{equation}
Substituting $\alpha=0$ into the contour integral of \eqref{diagfinal}
and using the representation \eqref{fNLIE} for the bulk free energy
density we get the relation
\begin{equation}
  \lim_{h_z\to\infty} (F_B+\frac{h_z}{T})=F_B(\alpha=\eta,\beta=\infty)-f.
\end{equation}
The right hand side describes the boundary free energy of
a chain with one less sites and longitudinal magnetic field
$h_z=\sinh(\eta)\coth(\eta)=\cosh(\eta)$. This is exactly what we
expect from the Hamiltonian \eqref{H} after we freeze one of the boundary
spins and correct for the diverging additive term.

\bigskip

An other interesting special case of \eqref{final} is when $h_z=0$, but $h_x$ is finite. This
is achieved by choosing $\beta=0$ such that
\begin{equation}
  h_x=\frac{ \sinh(\eta)}{\sinh(\alpha)}.
\end{equation}
For this case we obtain
\begin{equation}
  \label{finalhx}
  \begin{split}
 F_B/T=&
  \frac{1-\delta_{\Delta>1}}{2}\log(2)
-\delta_{-\alpha}\log(1+\fa(-\alpha))-
 \\
 & -\oint\limits_{\mc{C}}\f{\mathrm{d} \nu}{2\pi i}
 \left[\coth(2\nu)+\coth(2\nu+\eta)-\tanh(\nu)-\coth(\nu+\alpha)
   \right]\log(1+\fa(\nu))-\\
   &
   -\frac{1}{2}\log
 \frac{\det\left(1-\hat H^{+}\right)}{\det\left(1-\hat H^-\right)}
 +
 \frac{\coth(\alpha)\sinh(\eta)}{T}.
\end{split}
\end{equation}

The important case of the free boundary conditions ($h_x=h_z=0$) is
obtained by substituting $\alpha=i\pi/2$ into \eqref{diagfinal} or by
taking the $\alpha\to\infty$ limit of \eqref{finalhx}. In either case
we get
\begin{equation}
  \label{elegjo}
  \begin{split}
    F_B/T=& \frac{1-\delta_{\Delta>1}}{2}\log(2)
    -\frac{1}{2}\log
 \frac{\det\left(1-\hat H^{+}\right)}{\det\left(1-\hat H^-\right)}-
   \\
& -\oint\limits_{\mc{C}}\f{\mathrm{d} \nu}{2\pi i}
\left[\coth(2\nu)+\coth(2\nu+\eta)-\tanh(\nu)
     \right]  \log(1+\fa(\nu)).
\end{split}
\end{equation}

\subsection{Boundary magnetization}

Here we compute the expectation values of the boundary spin operators
using the relations
\begin{equation}
  \vev{\sigma_{1}^a}=\vev{\sigma_L^a}=
  \frac{\partial F_B}{\partial h^a},\qquad a=x,z.
\end{equation}
Applying the chain rule we get
\begin{equation}
  \label{finalmag}
  \begin{pmatrix}
    \vev{\sigma_1^z}& \vev{\sigma_1^x}
  \end{pmatrix}=
 \begin{pmatrix}
    \frac{\partial F_B}{\partial \alpha}&
      \frac{\partial F_B}{\partial \beta}
  \end{pmatrix}
  \begin{pmatrix}
    \frac{\partial\alpha}{\partial h_z} &
    \frac{\partial\alpha}{\partial h_x} \\
  \frac{\partial\beta}{\partial h_z} &
    \frac{\partial\beta}{\partial h_x} 
  \end{pmatrix}.
\end{equation}
The Jacobian is found to be
\begin{equation}
  \begin{split}
      \begin{pmatrix}
    \frac{\partial\alpha}{\partial h_z} &
    \frac{\partial\alpha}{\partial h_x} \\
  \frac{\partial\beta}{\partial h_z} &
    \frac{\partial\beta}{\partial h_x} 
  \end{pmatrix}=
    \frac{\sinh(\alpha)\cosh(\beta)}{\sinh(\eta)(\sinh^2(\beta)+\cosh^2(\alpha))}
       \begin{pmatrix}
     -\sinh(\beta)\sinh(\alpha)  &
     - \cosh(\alpha)\sinh(\alpha) \\
        \cosh(\alpha)\cosh(\beta)&
      -\sinh(\beta)\cosh(\beta)
    \end{pmatrix}, 
  \end{split}
\end{equation}
and the $\alpha,\beta$-derivatives are
\begin{equation}
  \label{finalmagneto}
  \begin{split}
 \frac{\partial F_B}{\partial \alpha}=&
 -\frac{1}{\sinh^2(\alpha)\sinh(\eta)}
+  T \delta_{-\alpha}\frac{\fa'(-\alpha)}{1+\fa(-\alpha)}
 -T\oint\limits_{\mc{C}}\f{\mathrm{d} \nu}{2\pi i}
 \frac{\log(1+\fa(\nu))}{\sinh^2(\nu+\alpha)}\\
  \frac{\partial F_B}{\partial \beta}=&
 \frac{1}{\cosh^2(\beta)\sinh(\eta)}
+  T \delta_{-\beta-i\pi/2}\frac{\fa'(-\beta-i\pi/2)}{1+\fa(-\beta-i\pi/2)}
 +T\oint\limits_{\mc{C}}\f{\mathrm{d} \nu}{2\pi i}
 \frac{\log(1+\fa(\nu))}{\cosh^2(\nu+\beta)}.\\
\end{split}
\end{equation}
If needed, the derivatives of the auxiliary function can be computed
using the NLIE \eqref{NLIE}.

\section{High temperature limit}

\label{sec:highT}

In this Section we compute the high temperature limit of the boundary
free energy. This provides a highly non-trivial check of the final
results of the previous section.

First we compute the high $T$ limit from the definition of the
partition function.
In any finite volume we have to first order in $1/T$
\begin{equation}
  Z=\text{Tr}\ e^{- H/T}=
\text{Tr}\ \mathbf{1}- \text{Tr} H/T+\ordo(T^{-2}).
\end{equation}
In the normalization \eqref{H} this gives
\begin{equation}
  Z=2^L(1+T^{-1}\Delta(L-1))+\ordo(T^{-2}),\qquad
  F=-TL\log 2-\Delta (L-1)+\ordo(T^{-1}).
\end{equation}
The bulk piece is $f= -T\log(2)-\Delta+\ordo(T^{-1})$ and we can read off the boundary contribution
\begin{equation}
  \label{hightfb}
  \lim_{T\to\infty} 2 F_B= \Delta.
\end{equation}
This constant is not physical, it merely represents the
additive normalization of the bulk term in the
Hamiltonian. Nevertheless we re-derive it from the general result
\eqref{final}.

For simplicity we consider the case when $\alpha,\beta$ lie outside
the contour, but the manipulations will be valid for arbitrary $\Delta$. In the high temperature limit we have
from \eqref{NLIE} $\lim_{T\to\infty}\fa(\lambda)=1$. The first
correction is easily found from \eqref{NLIE}. 
Defining
\begin{equation}
  \fa(\lambda)=1+T^{-1}G(\lambda)+\ordo(T^{-2})
\end{equation}
we get the linear integral equation
\begin{equation}
  \label{Gi}
G(\lambda)=
-\frac{2\sinh^2(\eta)}{\sinh(\lambda)\sinh(\lambda+\eta)}
+ \int_\CC \frac{d\nu}{2\pi i} 
\varphi(\lambda-\nu)
\frac{G(\nu)}{2}.
\end{equation}
The solution is
\begin{equation}
  \label{Gsol}
  G(\lambda)=\sinh(\eta)\left[
    \frac{\cosh(\lambda+\eta)}{\sinh(\lambda+\eta)}+
    \frac{\cosh(\lambda-\eta)}{\sinh(\lambda-\eta)}-2
       \frac{\cosh(\lambda)}{\sinh(\lambda)}
    \right].
\end{equation}
This is obtained iteratively from \eqref{Gi} using contour integral
manipulations. The solution \eqref{Gsol} can be checked by
substituting it back into \eqref{Gi} and noting that in the contour
integral the only singularity is a simple pole at $\nu=0$, whose
contribution is $-\sinh(\eta)\varphi(\lambda)$.

As a first step we prove that $\lim_{T\to\infty} F_B/T=0$. To this
order it is enough to substitute $\fa\equiv 1$ into
\eqref{final}. This gives immediately
\begin{equation}
  \label{final2}
  \begin{split}
\lim_{T\to\infty} F_B/T=&
-\lim_{T\to\infty}\frac{1}{2}\log
 \frac{\det\left(1-\hat H^{+}\right)}{\det\left(1-\hat H^-\right)}.
\end{split}
\end{equation}
The Fredholm determinants can be analyzed using the expansion \eqref{dorey-tateo}.
To lowest order in $T^{-1}$ the weight function in the kernels
\eqref{kernels} becomes $\fa/(1+\fa)=1/2$. Each trace is a multiple
contour integral over functions that are free of poles within $\CC$,
thus in this limit both determinants evaluate to unity, and we get
indeed $\lim_{T\to\infty} F_B/T=0$. 

To get the first corrections we use the approximations
\begin{equation}
  \begin{split}
  \log(1+\fa(\lambda))=\log(2)+\frac{1}{2T}G(\lambda)
  +\ordo(T^{-2})\qquad
  \frac{\fa(\lambda)}{1+\fa(\lambda)}=\frac{1}{2}+\frac{1}{4T}G(\lambda)
   +\ordo(T^{-2}).
   \end{split}
\end{equation}
In the expansion \eqref{dorey-tateo} only the term with $n=1$ contributes, leading to
\begin{equation}
  \begin{split}
    -  \frac{1}{2}\log
 \frac{\det\left(1-\hat H^{+}\right)}{\det\left(1-\hat H^-\right)}
&=\frac{1}{2}
\int_{\mathcal{C}} \frac{d\lambda}{2\pi i}
\frac{1}{4T}G(\lambda)
 \varphi(2\lambda)+\ordo(T^{-2})=\\
 &=-\frac{1}{2}\frac{1}{T} \cosh(\eta)+\ordo(T^{-2}).
 \end{split}
\end{equation}
It follows that
\begin{equation}
\begin{split}
 & \lim_{T\to\infty} F_B=-\frac{1}{2}\cosh(\eta)
+(\coth(\alpha)+\tanh(\beta))\sinh(\eta)-\\
& -\oint\limits_{\mc{C}}\f{\mathrm{d} \nu}{2\pi i}
 \left[\coth(2\nu)+\coth(2\nu+\eta)-\coth(\nu+\alpha)-\tanh(\nu+\beta)
 \right]
 \frac{G(\lambda)}{2}=\\
& -\frac{1}{2}\cosh(\eta)+\cosh(\eta)=\frac{1}{2}\cosh(\eta).
\end{split}
\end{equation}
This is in accordance with \eqref{hightfb}.

Higher order corrections in $1/T$ could be obtained by taking further
derivatives of the NLIE and thus by computing a number of simple contour integrals. The calculations
would be analogous to the high-$T$ expansion of the free energy itself
\cite{DdV2}. We do not pursue this direction here, because real
space calculations for the high temperature expansion are considerably
simpler. For the sake of completeness we give the second term:
\begin{equation}
\label{highT2}
  F_B=\frac{\Delta}{2}+
 \frac{1}{T}\left[\frac{2+\Delta^2}{4}-\frac{h_x^2+h_z^2}{2}\right]+\ordo(T^{-2}).
\end{equation}

\section{The $T\to 0$ limit}

\label{sec:T0}

Here we investigate the low temperature behaviour of the boundary
free energy.

In the $T\to 0$ limit $F_B$ 
becomes the boundary energy $E_B$ of the ground state, which is defined as
\begin{equation*}
2 E_B=\lim_{L\to\infty}  (E_0(L)-e_0L),
\end{equation*}
where $E_0(L)$ is  the finite volume ground state energy and $e_0$ is the
ground state energy density. Once again we
assumed identical boundary conditions at the two ends of the chain.

For $E_B$ a number of results have been already
derived, both for purely longitudinal \cite{skorik-saleur,skorik-kapustin}
and arbitrary boundary fields
\cite{nepomechie-elbaszott-boundary,kinai-trukkos-boundary,kinai-su3-boundary,off-diagonal-book}.
The final results depend on the root content of the ground
state: depending on the parameters there can be boundary bound
states, that are represented by special Bethe roots. Therefore, the
discussion of the boundary energy has to be performed carefully, as
it was done in the works cited above.

In this section we compute $E_B$ in both regimes and reproduce
earlier results in the literature. In the massive regime we also investigate the first low-$T$
correction terms: we argue that in most cases there is linear term in
$T$ whose coefficient is exactly obtained.  
In the massless case we discuss the earlier predictions of QFT methods.

\subsection{The massive regime}

It is useful to introduce the $\fb$-$\bar\fb$
formulation \cite{kluemper-QTM,kluemper-review} of the NLIE. First
we make use of the coordinate transformation $\lambda=ix$. We define
$\tilde\fa(x)=\fa(ix)$.
The integral equation becomes
\begin{equation}
\begin{split}
  \log \tilde \fa(u)=
2\beta \frac{\sinh\eta}{\sin(u)\sin(u-i\eta)}
-\int_C\frac{d\nu}{2\pi}
\frac{\sinh(2\eta)}{\sin(u-\nu+i\eta)\sin(u-\nu-i\eta)}
 \log
(1+\tilde \fa(\nu)).
\end{split}
\end{equation}
We define the functions
\begin{equation}
  \fb(x)=\lim_{\eps\to 0}\tilde \fa(x-i\eta/2+i\eps)
 \qquad\bar\fb(x)=\lim_{\eps\to 0}\tilde \fa^{-1}(x+i\eta/2-i\eps)
\end{equation}
and
\begin{equation}
  \label{fB}
  \fB(x)=1+ \fb(x)\qquad
  \bar\fB(x)=1+\bar \fb(x).
 \end{equation}
 These functions satisfy the symmetry properties
\begin{equation}
 \bar\fb(x)=\fb(-x)=\fb^*(x)\qquad
  \bar\fB(x)=\fB(-x)=\fB^*(x).
\end{equation}
The NLIE can be transformed to the form (for a detailed derivation see \cite{aa-to-bb})
\begin{equation}
  \label{bb}
\begin{split}
&  \log \fb(x)=-\frac{2\sinh(\eta)}{T} s(x)
+\int \frac{dy}{2\pi}\kappa(x-y) \log \fB(y)-\\
&\hspace{4cm}-\lim_{\eps\to 0}\int \frac{dy}{2\pi} \kappa(x-y-2i(\eta-\eps)) \log \bar\fB(y),
\end{split}
\end{equation}
where
\begin{equation}
\label{sdef}
s(x)=1+2\sum_{n=1}^\infty \frac{\cos(2n x)}{\cosh(\eta n)},
\end{equation}
and
\begin{equation}
  \kappa(x)=\sum_{k=-\infty}^\infty \frac{e^{-|k|\eta}}{\cosh(k\eta)}e^{2ik x}.
\end{equation}
Note that $\kappa$ is well defined only if $|\Im(x)|<\eta$ and it has
a pole at $x=\pm i\kappa$.
This is the reason why this set of auxiliary functions and the NLIE
are only defined
through the $\eps\to 0$ limit.

It follows from the NLIE that in the $T\to 0$ limit $\tilde\fa(x)$
becomes exponentially large/small if $x$ is in the upper/lower
half-plain. For $x\in\valos$ we have $|\tilde\fa(x)|=1$ for
arbitrary temperatures (this follows from the original definition
\eqref{adef} and the symmetry of the Bethe roots), nevertheless the
function $\tilde\fa(x)$ does not have a finite $T\to 0$ limit. An
exception is given by the special point $x=\pi/2$ for which we have
$\tilde\fa(\pi/2)=\fa(i\pi/2)=1$ at arbitrary temperatures.

For the limiting behaviour of the $\fb$-$\bar\fb$ functions we have 
\begin{equation}
\lim_{T\to 0} \Big( T\log(\fb(x))\Big)=\lim_{T\to 0}\Big(  T\log(\bar\fb(x))\Big)=
   -2\sinh(\eta) s(x).
 \end{equation}
Therefore the original contour integrals over any function $f(u)$ have
the limiting behaviour
\begin{equation}
  \label{fuint}
  \begin{split}
& \lim_{T\to 0} \left[ T\int_C \frac{du}{2\pi i} f(u) \log(1+\fa(u))\right]=
2\sinh(\eta) \lim_{\eps\to 0} \left[ \int \frac{dx}{2\pi}
 s(x-i\eps) f(ix-\eta/2+\eps)\right].
 \end{split}
\end{equation}
We are now in the position to compute the $T\to 0$ limit of the
formula \eqref{final}. It can be seen that the Fredholm determinants remain bounded,
because their weight functions converge to finite values (the limiting
value of the ratio will be computed at the end of this subsection). Therefore
the $T\to 0$ limit of the free energy is determined by the constants
and the simple integrals.

First we consider the case when there are no boundary bound states,
that is, when $\alpha$ and $\beta$
lie outside the physical strip. 
Using \eqref{fuint} we get
\begin{equation}
  \begin{split}
    \lim_{T\to 0} F_B&=
\sinh(\eta)\Big[\coth(\alpha)+\tanh(\beta)
+    2 \lim_{\eps\to 0} 
       \int \frac{dx}{2\pi}
       s(x-i\eps)
       \left[\coth(2ix-\eta+2\eps)+\right. \\
& \left.        +\coth(2ix+2\eps)-\coth(ix-\eta/2+\alpha+\eps)-\tanh(ix-\eta/2+\beta+\eps)
     \right] \Big].
     \end{split}  
   \end{equation}
The integrals can be evaluated using the following identities, that
hold for an arbitrary $\gamma\in\valos$:
\begin{equation}
  \label{ids}
  \begin{split}
  \int_{-\pi/2}^{\pi/2} \frac{dx}{2\pi} s(x)
  \coth(ix+\gamma)&=\text{sign}(\gamma)\sum_{n=-\infty}^{\infty}
  \frac{e^{-2|n\gamma|}}{2\cosh(n\eta)}\\
   \int_{-\pi/2}^{\pi/2} \frac{dx}{2\pi} s(x)
  \coth(2ix+\gamma)&=\text{sign}(\gamma)\sum_{n=-\infty}^{\infty}
  \frac{(1+(-1)^n)e^{-|n\gamma|}}{4\cosh(n\eta)}\\
    \int_{-\pi/2}^{\pi/2} \frac{dx}{2\pi} s(x)
  \tanh(ix+\gamma)&=\text{sign}(\gamma)\sum_{n=-\infty}^{\infty}
  \frac{(-1)^ne^{-2|n\gamma|}}{2\cosh(n\eta)}.\\
\end{split}  
\end{equation}
Putting everything together we obtain the following formula for the
boundary energy:
\begin{equation}
  \label{DE12}
 E_B=\lim_{T\to 0} F_B=\sinh(\eta)(e_0+e_\alpha+e_\beta),
\end{equation}
with
\begin{equation}
  \label{e0def}
    e_0=\frac{1}{2}\sum_{n=-\infty}^\infty
    \frac{(1-e^{-|n|\eta})(1+(-1)^n)} {\cosh(n\eta)}=
2 \sum_{n=1}^\infty   \frac{1-e^{-2n\eta}} {\cosh(2n\eta)}
\end{equation}
and
\begin{equation}
  \begin{split}
e_\alpha&=\coth(\alpha)-\text{sign}(\alpha-\eta/2)\left[1+
  2\sum_{n=1}^\infty
\frac{ e^{-|2\alpha-\eta|n}}{\cosh(n\eta)}\right]\\
e_\beta&=\tanh(\beta)-\text{sign}(\beta-\eta/2)
\left[1+2\sum_{n=1}^\infty
\frac{  (-1)^n  e^{-|2\beta-\eta|n}}{\cosh(n\eta)}\right].
\end{split}
\end{equation}
Expression \eqref{DE12} agrees with the earlier result for this
case published in \cite{off-diagonal-book}: it coincides with formula
 (5.4.47) after correcting for the differences  in the definition of
 the Hamiltonian, including certain signs of the boundary parameters. 

The case with boundary bound states can be treated similarly. For
example consider the case when $\alpha$ is within the physical
strip, and we need to evaluate the extra term
$T\log(1+\fa(-\alpha))=T\log(1+\tilde\fa(-i\alpha))$. It can be seen
from \eqref{bb} that if $\alpha<0$ then this term becomes
exponentially small, but it gives a finite contribution when
$\alpha>0$. In the latter case the NLIE yields
\begin{equation}
  \label{extra}
  \begin{split}
&  \lim_{T\to 0} \left[T \log (1+\fa(-\alpha))\right]=
\lim_{T\to 0} \left[T \log \fa(-\alpha)\right]=
  - \lim_{T\to 0} \left[T \log \tilde\fb(i(\alpha-\eta/2))\right]=
  \\
  &=  2\sinh(\eta) s(i(\alpha-\eta/2))=
  2\sinh(\eta)\left(  1+2\sum_{n=1}^\infty \frac{\cosh(n\eta-2n \alpha)}{\cosh(\eta n)}\right).
\end{split}
\end{equation}
This term needs to be added to expression \eqref{DE12} if
$0<\alpha<\eta/2$. A similar contribution has to be added when $0<\beta<\eta/2$.

An important test of the results is that the boundary energy should be
invariant with respect to the spin flips $h_z\to -h_z$ and/or $h_x\to
-h_x$, which corresponds to $\alpha,\beta\to -\alpha,-\beta$.
A simple calculation shows that if $\alpha$ and $\beta$ lie outside the
physical strip, then the functions 
$e_1(\alpha)$ and $e_2(\beta)$ are indeed symmetric. On the other
hand, if for example $|\alpha|<\eta/2$ then $E_B$ becomes symmetric
after we add the extra contribution \eqref{extra} when $\alpha>0$. 
The total boundary energy is thus always symmetric.

In view of these symmetry considerations the boundary energy is
always described by formula \eqref{DE12} with $e_0$ given by
\eqref{e0def} but $e_{\alpha,\beta}$ replaced by
\begin{equation}
  \label{eab}
  \begin{split}
e_\alpha&=-\coth(|\alpha|)+1+
  2\sum_{n=1}^\infty
\frac{ e^{-(2|\alpha|+\eta)n}}{\cosh(n\eta)}\\
e_\beta&=-\tanh(|\beta|)+
1+2\sum_{n=1}^\infty
\frac{  (-1)^n  e^{-(2|\beta|+\eta)|n}}{\cosh(n\eta)}.
\end{split}
\end{equation}

The specific case of free boundary conditions is obtained by 
evaluating the low-$T$ limit of the integrals in 
\eqref{elegjo}, or simply taking the $\alpha\to\infty$, $\beta\to
0$ limit of  \eqref{DE12}. Either way we get
\begin{equation}
  \label{Ebfree}
     \begin{split}
       E_B=
              \sinh\eta\left[1+2
         \sum_{n=1}^{\infty}\frac{
           (1-e^{-2n\eta})}
                {\cosh(2n\eta)}+
   2\sum_{n=1}^{\infty}\frac{(-1)^ne^{-n\eta}}{\cosh(n\eta)}
             \right].
   \end{split}     
 \end{equation}
After correcting for the differences in the definition of the
Hamiltonian this coincides with the formula (32) presented in
\cite{skorik-kapustin}.

\bigskip

It is also possible to compute the first low-$T$ correction to
$E_B$. Again, we first consider the case when $\alpha,\beta$ lie
outside the physical strip.
The NLIE \eqref{bb} implies that for very low temperatures the
correction terms to the simple integrals will be exponentially
small. Thus the leading correction is a linear term given by
\begin{equation}
F_B=E_B+T\left(\log(2) -
 \frac{1}{2}\lim_{T\to 0}  \log
 \frac{\det\left(1-\hat H^{+}\right)}{\det\left(1-\hat H^-\right)}
\right)+\dots.
\end{equation}
The limit of the Fredholm determinants is computed as follows. The
weight function $\tilde\fa(\nu)/(1+\tilde\fa(\nu)$ becomes
exponentially small on the upper contour, therefore it is enough to
consider the integrations over the lower contour on which
$\tilde\fa(\nu)/(1+\tilde\fa(\nu))\to 1$. We get
\begin{equation}
   \frac{\det\left(1-\hat H^{+}\right)}{\det\left(1-\hat H^-\right)}\quad\to\quad
\frac{\det\left(1-\hat H_0^{+}\right)}{\det\left(1-\hat H_0^-\right)},
 \end{equation}
where $H_0^{\pm}$ are integral operators that act on functions defined
on the segment $[-\pi/2\dots \pi/2]-ia$, $a<\eta/2$ as
\begin{equation}
  \label{kernels2}
  \begin{split}
    \left(\hat H_0^{\pm} (g)\right)(x)&=-\int_{-\pi/2}^{\pi/2} \frac{du}{2\pi}
 \frac{\varphi(x-u)\pm\varphi(x+u-2ia)}{2} g(u)=\\
&=-\int_{-\pi/2}^{\pi/2} \frac{du}{2\pi}
 \varphi(x-u)\frac{g(u)\pm g(-u-2ia)}{2}.
\end{split}
\end{equation}
The overall minus sign in front of the integrals originates in the
transition from the contour integrals to the real integrals.

The determinants can be evaluated in Fourier space. The integral
operator couples the Fourier modes $g_n$ and $g_{-n}$ for any $n>1$,
the matrices $1-\hat H_0^{\pm}$ are thus block-diagonal with $2\times 2$
blocks, except for the zero Fourier mode. The sign difference
between the the two operators affects the two off-diagonal matrix
elements in the $2\times 2$  blocks, therefore all sub-determinants
are actually equal. The difference between the two 
determinants is thus only in the zero mode, and we get
\begin{equation}
  \frac{\det\left(1-\hat H_0^{+}\right)}{\det\left(1-\hat H_0^-\right)}
  =\frac{1+\int_{-\pi/2}^{\pi/2} \frac{du}{2\pi} \frac{\varphi(x-u)+\varphi(x+u-2ia)}{2}}{1+\int_{-\pi/2}^{\pi/2} \frac{du}{2\pi}
  \frac{\varphi(x-u)-\varphi(x+u-2ia)}{2}}=
1+\int_{-\pi/2}^{\pi/2} \frac{du}{2\pi} \varphi(u)=2.
\end{equation}
The leading term to the boundary free energy is thus
\begin{equation}
  F_B=E_B+T\frac{\log(2)}{2}+\dots.
\end{equation}

In those cases when $\alpha$ or $\beta$ lie within the physical strip
we need to consider the low-$T$ corrections to the extra terms 
$\log(1+\fa(-\alpha))$ and
$\log(1+\fa(-\beta-i\pi/2))$. However, the NLIE
implies that the corrections will be exponentially small, except the
special case of $\log(1+\fa(-i\pi/2))=\log(2)$ which holds for
arbitrary temperatures. This is encountered only when $h_z=0$ and
according to \eqref{finalhx} it results in
\begin{equation}
  F_B=E_B-T\frac{\log(2)}{2}+\dots.
\end{equation}
In the massive regime the $\ordo(T)$ terms can be summarized as
\begin{equation}
  \label{massiveT0}
   F_B=E_B+T\log(2)\left(\frac{1}{2}-\delta_{h_z=0}\right)+\dots,
\end{equation}
where the dots denote corrections that are exponentially suppressed in $1/T$.

\subsection{The massless regime}

\label{sec:masslessT0}

Now $\eta=i\gamma$ with $\gamma\in\valos$. We will also use
$\alpha=i\tilde\alpha$ such that $\tilde\alpha\in\valos$.

The $\fb-\bar \fb$
formulation of the NLIE is introduced as follows. We define
\begin{equation}
    \fb(x)=\lim_{\eps\to 0} \fa(x+i\gamma/2+i\eps)
   \qquad\bar\fb(x)=\lim_{\eps\to 0} \fa^{-1}(x-i\gamma/2-i\eps),
\end{equation}
and $\fB-\bar\fB$ are given by \eqref{fB}.
The NLIE can be transformed into the form  \cite{kluemper-QTM,aa-to-bb}
\begin{equation}
  \label{bbmassless}
  \begin{split}
  \log \fb(x)&=-\frac{2\sin(\gamma)}{T} d(x)
+\int \frac{dy}{2\pi}\kappa(x-y) \log \fB(y)-\\
&\hspace{3cm}-\lim_{\eps\to 0}\int \frac{dy}{2\pi} \kappa(x-y-2i(\eta-\eps)) \log \bar\fB(y),
\end{split}
\end{equation}
where now
\begin{equation}
  d(x)=\frac{\pi}{\gamma \cosh(\pi x/\gamma)}
\end{equation}
and
\begin{equation}
  \kappa(x)=\int_{-\infty}^\infty \frac{dk}{2\pi}
  \frac{ \sinh((\pi/2-\gamma)k) }{2\cosh(\gamma
    k/2)\sinh((\pi/2-\gamma/2)k)} e^{ikx}.
\end{equation}
The asymptotic behaviour of the auxiliary functions is similar to that
of the massive case, nevertheless there are important differences that
result in other types of correction terms. The NLIE implies that in
the low-$T$ limit $\fa(x)$ becomes exponentially large/small for $x$
in the lower/upper half-plains, respectively. However, for any fixed
$T$ the asymptotics is always given by
\begin{equation}
  \lim_{\Re(x)\to \pm\infty} \fa(x)=1.
\end{equation}
For any finite $T$ there is a crossover regime in $x$ where the source
term of the NLIE \eqref{bbmassless} is $\ordo(1)$, this happens when
\begin{equation}
  \label{x0def}
|x|\approx x_0=\frac{\gamma}{\pi}\log\left[ \frac{4\pi\sin(\gamma)}{T\gamma}\right].
\end{equation}
In this regime a $T$-independent asymptotic NLIE can be written down for a shifted
rapidity variable $x-x_0$, whose solution describes a kink that moves
with $T$ in rapidity space  \cite{DdV2}. The logarithmic dependence
implies that depending the problem at hand the temperature
must be chosen extremely low to reach the asymptotic regime.  This
will have a relevance to our numerical data for $F_B$.

Now we compute the boundary energy in this regime. Once again the
$T\to 0$ limit of $F_B$ will be
given by the simple integrals and the constant terms, because the
Fredholm determinants will remain regular and thus they only contribute an
$o(1)$ term to $F_B$. 
The  boundary parameter $\beta$ is real and thus in this regime the
special point $-\beta-i\pi/2$ is always outside the canonical
contour. Therefore
\begin{equation}
  \label{finalkkk}
  \begin{split}
& E_B=\lim_{T\to 0} F_B=
 -\lim_{T\to 0}\delta_{-\alpha}\log(1+\fa(-\alpha))   +i(\coth(\alpha)+\tanh(\beta))\sin(\gamma)-
 \\
 & -\lim_{T\to 0}\oint\limits_{\mc{C}}\f{\mathrm{d} \nu}{2\pi i}
 \left[\coth(2\nu)+\coth(2\nu+\eta)-\coth(\nu+\alpha)-\tanh(\nu+\beta)
   \right]\log(1+\fa(\nu)).
   \end{split}
\end{equation}
In the following we concentrate on the case when $\alpha$ is outside
the physical strip. It follows from the NLIE that in the $T\to 0$
limit the leading contribution comes from the lower contour and it
reads
\begin{equation}
  \label{finalkk2}
  \begin{split}
    E_B=&   +i(\coth(\alpha)+\tanh(\beta))\sin(\gamma)- \lim_{\eps\to 0}\int_{-\infty}^\infty \f{\mathrm{d} \nu}{2\pi i}
    \left[\coth(2\nu-i\gamma)+\right.\\
&\left.
  +\coth(2\nu+2i\eps)    -\coth(\nu+\alpha-i\gamma/2)-\tanh(\nu+\beta-i\gamma/2)
  \right] 2\sin(\gamma) d(x).
   \end{split}
\end{equation}
After symmetrizing the integral and using
\begin{equation}
  \lim_{\eps\to 0}(\coth(2\nu+2i\eps)-\coth(2\nu-2i\eps))
  =-\pi\delta(\nu)  
\end{equation}
we get
\begin{equation}
  \label{finalkk2b}
  \begin{split}
  &  E_B=
    \sin(\gamma)\left\{i(\coth(\alpha)+\tanh(\beta))+\frac{d(0)}{2}
    -\int_{-\infty}^\infty \f{\mathrm{d} x}{2\pi i}
\left[
  \frac{\sin(2\gamma)}{\sinh(2x-i\gamma)\sinh(2x+i\gamma)}-\right.\right.\\
&\left.\left.      +\frac{\sinh(2\alpha-i\gamma)}{\sinh(x+\alpha-i\gamma/2)\sinh(x-\alpha+i\gamma/2)}
-\frac{\sinh(2\beta-i\gamma)}{\cosh(x+\beta-i\gamma/2)\cosh(x-\beta+i\gamma/2)}
  \right] d(x)\right\}.
   \end{split}
\end{equation}
The case of free boundary conditions is obtained by setting
$\alpha=i\pi/2$ and performing the limit $\beta\to\infty$. In this
case we obtain
\begin{equation}
  \label{finalkk3}
  \begin{split}
    E_B=&  \sin(\gamma)\frac{d(0)}{2}
- \sin(\gamma)\int_{-\infty}^\infty \f{\mathrm{d} x}{2\pi }
\left[
  \frac{\sin(2\gamma)}{\sinh(2x-i\gamma)\sinh(2x+i\gamma)}+\right.\\
&\hspace{1cm}  \left.    +\frac{\sin(\gamma)}{\cosh(x-i\gamma/2)\cosh(x+i\gamma/2)}
  \right]  d(x).
   \end{split}
 \end{equation}

It is an important task to consider the first low-$T$ corrections in
$F_B$ and to compare to predictions from QFT.
The works
\cite{Fujimoto:PhysRevLett92:2004,Bortz:JPhysAMathGen38:2005,Affleck-XXZ-boundary-QFT} 
treated the case of free boundary conditions using Conformal Field
Theory (CFT) and bosonization techniques.
These papers concentrated on the boundary contribution to the total
susceptibility, therefore they computed $F_B$ in the general case with
a finite $h$ bulk magnetic field. Their formulas should
describe the low-$T$ behaviour of our results for $h=0$. Correcting
for the differences in the definitions of the Hamiltonian and $F_B$ their result reads
\begin{equation}
  \label{masslessQFT}
  F_B=E_B+\frac{\lambda}{2} \frac{\Gamma(K)\Gamma(1-2K)}{\Gamma(1-K)}
\left(\frac{2\pi T}{v}\right)^{2K-1}+\ordo(T^2),
\end{equation}
where
\begin{equation}
  \begin{split}
  K&=\frac{1}{1-\gamma/\pi}\\
  v&=\frac{2K\sin(\pi/K)}{K-1}\\
  \lambda&=\frac{4K\Gamma(K)\sin(\pi/K)}{\pi \Gamma(2-K)}
  \left(\frac{\Gamma(1+\frac{1}{2K-2})}{2\sqrt{\pi}\Gamma(1+\frac{K}{2K-2})}\right)^{2K-2}.
  \end{split}
\end{equation}
Most importantly they found that there is no linear term in $F_B$.

At present we are not able to confirm or disprove the above
prediction. The behaviour of the integrals and Fredholm-determinants
is considerably more involved in the massless regime than in the
massive. On the one
hand, if we choose
$\CC$ to be a compact contour then $\CC$ needs to be changed as we
lower the temperature. On the other hand, if we choose $\CC$ to be the
union of two straight lines, then both the single integrals and the
Fredholm determinants become ill behaving. We plan to return to this
problem in future research.

\section{The XXX limit}

\label{sec:XXX}

Here we compute the boundary free energy at the isotropic point. In
principle we could repeat the whole calculation using the $R$ and $K$
matrices of the XXX model, but it is easier to take the limits
of the final formulas. The boundary free energy is a continuous
function of $\Delta$, and we choose to approach $\Delta=1$ from the massless
regime.

It is known that with the parametrization $\Delta=\cos(\gamma)$ the
$\gamma\to 0$ limit leads to a re-scaling of the rapidity variables according
to
\begin{equation}
  \lambda_{XXZ}\quad\to \gamma \lambda_{XXX},
\end{equation}
such that the $\lambda_{XXX}$ become the finite variables of the XXX
model.

After this re-scaling the NLIE \eqref{NLIE} takes the form  
\begin{equation}
\label{NLIExxx}
\begin{split}
\log \fa(\lambda)=&
\frac{1}{T}\frac{2}{\lambda(\lambda+i)}
+  \int_\CC \frac{d\nu}{2\pi} 
\tilde \varphi(\lambda-\nu) \log(1+\fa(\nu)),
\end{split}
\end{equation}
where
\begin{equation}
\tilde  \varphi(u)=-\frac{2}{u^2+1}
\end{equation}
and the contour $\CC$ encircles the real line such that it remains in
the physical strip $|\Im(\lambda)|<1/2$. In practice $\CC$ can be
chosen as the union of $\valos-i\kappa$ and $\valos+i\kappa$ with $\kappa<1/2$.

The re-scaling of the boundary parameters can be read off
\eqref{hdef}. Finite fields are obtained with the
re-scaling $\alpha\to i\gamma\tilde\alpha$ and keeping $\beta$ finite,
such that in the XXX limit
\begin{equation}
  h_x=\frac{1}{\tilde\alpha\cosh(\beta)}\qquad
  h_z=\frac{1}{\tilde\alpha} \tanh(\beta),
\end{equation}
where $\tilde\alpha,\beta\in\valos$. It can be seen that the amplitude
of the magnetic field is $1/|\tilde\alpha|$ and the $\beta$ parameter
only describes an angle in the $z-x$ plain.

Performing the scaling on \eqref{final} we obtain the boundary free
energy as
\begin{equation}
  \label{finalxxx}
  \begin{split}
 F_B/T=&
  \frac{1}{2}\log(2)
  -\oint\limits_{\mc{C}}\f{\mathrm{d} \nu}{2\pi i}
 \left[\frac{1}{2\nu}+\frac{1}{2\nu+i}-\frac{1}{\nu+i\tilde\alpha}\right]
 \log(1+\fa(\nu))-\\
   &
 -\delta_{-i\tilde\alpha}\log(1+\fa(-i\tilde\alpha))   -\frac{1}{2}\log
    \frac{\det\left(1-\hat H_{XXX}^{+}\right)}{\det\left(1-\hat H_{XXX}^-\right)}
 +
\frac{1}{\tilde\alpha T},
\end{split}
\end{equation}
where $\hat H_{XXX}^{\pm}$
are integral operators that
act as
\begin{equation}
  \label{kernelsxxx}
  \begin{split}
    \left(\hat H_{XXX}^{\pm} (g)\right)(x)&=\int_{\mathcal{C}}
    \frac{du}{2\pi } \frac{\fa(u)}{1+\fa(u)}
    \frac{\tilde\varphi(x-u)\pm\tilde \varphi(x+u)}{2} g(u).
\end{split}
\end{equation}
The ratio of Fredholm determinants is given alternatively by 
\eqref{dorey-tateo} with the replacements
\begin{equation}
  \int \frac{d\lambda_j}{2\pi i}\varphi(\dots) \qquad\to\qquad
    \int \frac{d\lambda_j}{2\pi }\tilde\varphi(\dots)
\end{equation}

It is an important consistency check of the scaling procedure
that the $\beta$-dependent contributions vanished, signaling the
$SU(2)$-invariance of the problem. 

Free boundary conditions are obtained in the $\tilde\alpha\to \infty$
limit, when all $\tilde\alpha$-dependent terms disappear from \eqref{finalxxx}.

\section{The ferromagnetic chain}

\label{sec:ferro}

The parameter choice $\Delta<-1$ describes the ferromagnetic spin chain, and
here we explain how the previous formulas can be used
for $\Delta<0$ as well.

Let us consider the total free energy in any finite volume of even length,
and let us denote it by $F(\Delta,T)$. 
Performing
a similarity transformation with the operator
$U=\prod_{j=1}^{L/2}\sigma^z_{2j-1}$
we get the relation \cite{Takahashi-book}:
\begin{equation}
  F(\Delta,T)=F(-\Delta,-T).
\end{equation}
Negative values of $\Delta$ can thus be treated by using negative
values of $T$ in our formalism. The main equations
 \eqref{NLIE} and \eqref{final} are analytic in the parameter $1/T$,
 therefore they can be continued naturally to the negative
 regime. This was already demonstrated in
 \cite{kluemper-goehmann-finiteT-review}. 

Naturally, the low-$T$ behaviour of $F_B$ is quite different in
the ferromagnetic regime. In the case of free boundary conditions the
two ground states are the
ferromagnetic reference states, such that $E_B=\lim_{T\to 0} F_B=0$. In the case of finite boundary fields
there can be boundary bound states, that give a finite contribution to
$E_B$.

These results can be obtained from our formulas; the low-$T$ behaviour
of the NLIE \eqref{NLIE} can be studied similarly to the
anti-ferromagnetic case. A $\fb$-$\bar\fb$ formulation can be written
down similar to \eqref{bb}, but the source terms and
integration kernels will be different.
We omit the details of these calculations, because the main focus of
this work is the anti-ferromagnetic chain.

\section{Numerical investigations}

\label{sec:numerics}

In this section we numerically evaluate the exact formulas for
$F_B$. We only set a modest goal of giving a few examples
of the numerical data; the
complete exploration of the behaviour of $F_B$ as a function of the
temperature, the anisotropy, and the two boundary fields requires a
separate study.

The NLIE \eqref{NLIE} can be implemented easily in both the
massive and massless regimes. An important point is the choice of the
contour $\CC$. For $\Delta>1$ we chose it to be the union of the
segments  $[-i\pi/2+a\dots i\pi/2+a]$ and
$[-i\pi/2-a\dots i\pi/2-a]$ with $a<\eta/2$. In the massless regime
the NLIE can be solved on the usual contour which is the union of
$\valos+ia$ and $\valos-ia$ with $a<\cos^{-1}(\Delta)/2$. In our
implementation we used the original $\fa$-formulation and the NLIE
could be solved with iteration up to low temperatures of order
$T=10^{-2}$. An important point is that the function
$\log(1+\fa(\lambda))$  has to be continuous in $\lambda$, and
artificial jumps of $2\pi i$ produced by the numerics have to be
corrected. 

Having found the numerical solution for $\fa(\lambda)$ the integrals
and the Fredholm determinants in \eqref{final} can be evaluated
too. The determinants can be calculated by discretizing the integral
operators and taking the determinant of the resulting matrix; this
can be worked out even for curved contours. This is in fact required in the massless
regime, where the expressions in \eqref{final} have weak convergence
properties if $\CC$ is the union of the two straight lines as
explained above. Instead, we considered an elliptic curve centered at
$\lambda=0$, where the major axis is large enough to include all Bethe
roots of the QTM, and the minor axis is small enough so that the
ellipse fits into the physical strip. This way the numerical data
became convergent and stable with respect to changing the parameters
of the ellipse.

As a first test of our results we compared them to exact diagonalization.
We computed the partition function
\eqref{eredetiZ} on a finite chain and computed the boundary
contribution as
\begin{equation*}
 2 F_{B}(L)=F_{ED}(L)-fL,
\end{equation*}
where $f$ is the exact value of the free energy density calculated
from the NLIE \eqref{NLIE}.
We performed this procedure on finite chains up to $L=12$ and $L=13$,
and for various values of the parameters $T$ and magnetic fields.
It was observed that $F_B(L)$ converged well for high or intermediate
temperatures, when the correlation length in the system is relatively
small. On the other hand, convergence was worse for lower
temperatures, especially in the massless regime, as it was
expected. We also observed that the convergence depends strongly on
the values of the magnetic fields.
Examples for the numerical data are shown in Tables \ref{tab:dat1} and
\ref{tab:dat2}.

\begin{table}[!h]
  \centering
  \begin{tabular}{|c|c||c||c|c|}
    \hline
    $\Delta$ & $T$ & $F_B$ & $L=12$ & $L=13$ \\
    \hline
    \hline
    0.5  &  0.1  &  1.20833  &  1.15943  &  1.27526 \\
        \hline
0.5  &  0.5  &  1.22386  &  1.22233  &  1.2248 \\ 
    \hline
    0.5  &  1  &  1.17986  &  1.17986  &  1.17986 \\ 
 \hline
    0.5  &  2  &  1.00888  &  1.00888  &  1.00888 \\ 
    \hline
    \hline
    2  &  0.1  &  3.52883  &  3.39913  &  3.55386 \\ 
 \hline
2  &  1  &  3.25907  &  3.25874  &  3.25923 \\ 
    \hline
    2  &  2  &  3.08094  &  3.08094  &  3.08094 \\ 
 \hline
    \hline
    -2 & 0.5 & 0.17376   & 0.17376  & 0.17376 \\
    \hline
    -2 & 1 & 0.34185 & 0.34185 & 0.34185  \\
    \hline
-2 & 2 &  0.54549 &0.54549  & 0.54549 \\     
    \hline
  \end{tabular}
  \caption{The exact boundary free energy at different anisotropies and
    temperatures, compared to results from exact
    diagonalization. The data refer to free boundary conditions.
}
 \label{tab:dat1}
\end{table}

\begin{table}[!h]
  \centering
  \begin{tabular}{|c|c||c||c|c|}
    \hline
    $\Delta$ & $T$ & $F_B$ & $L=12$ & $L=13$ \\
    \hline
    \hline
    2  &  1  &  0.46150  & 0.59893   & 0.36490  \\ 
    \hline
    2  &  2  & 0.64993  &  0.66270   &  0.64094  \\
    \hline
    2 &   5 &   0.83512 &    0.83513  &  0.83512  \\
    \hline    
  \end{tabular}
  \caption{Examples for the exact boundary free energy with off-diagonal magnetic
    fields. Here  $h_z=2$, $h_x=1$.
}
 \label{tab:dat2}
\end{table}

Whereas exact diagonalization confirmed our data for intermediate and
large temperatures, there were always sizable differences for small
temperatures due to the large finite size effects. In order to test
our formulas at low-$T$ we also compared them to unpublished numerical results
obtained using DMRG, in the case of free boundary conditions. This
data was produced and sent to us by Jesko 
Sirker, for which we are very grateful. The DMRG data confirmed
the correctness of our formulas down to temperatures of order
$T\approx 10^{-2}$. 

In order to explore the temperature dependence of
$F_B$ we produced numerical data for specific values of the anisotropy and
the boundary magnetic fields. We found that as a function of $T$ there can be a variety of
behaviors.

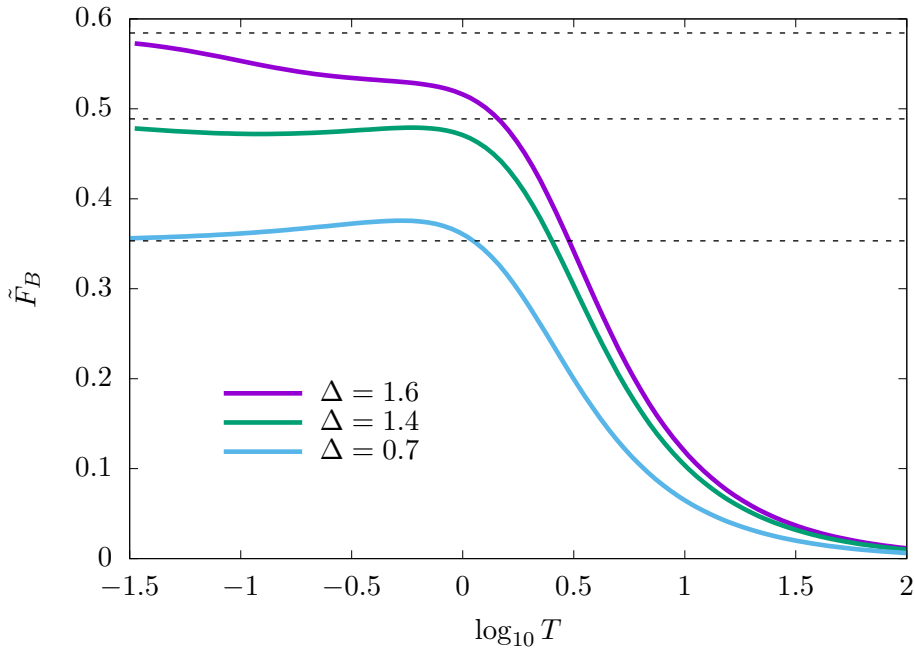
\begin{figure}
  \centering
  % AbraKozos.plt
\input{AbraKozos}
  \caption{The boundary free energy in the case of $h_z=h_x=0$  as a function of the temperature
    for different values of the anisotropy. We plotted $\tilde F_B=F_B-\Delta/2$
    such that all curves tend to 0 in the high temperature limit. The
    horizontal dashed lines show the $T\to 0$ limit as calculated from
    \eqref{Ebfree} and \eqref{finalkk3}.}
  \label{fig:kozos1}
\end{figure}

In the case of free boundary conditions three examples are shown in Fig \ref{fig:kozos1}. Here we plot
$\tilde F_B\equiv F_B-\Delta/2$ for three different values of
$\Delta$. The shift of $-\Delta/2$ was added so that all curves approach 0
in the high temperature limit; $\tilde F_B$ corresponds to the
boundary free energy of the Hamiltonian without the additive term in
the bulk piece. We found that for large anisotropy $\tilde F_B$ is a monotonically
decreasing function, but for smaller $\Delta$ it develops a local
minimum and a local maximum, see for example the curve with
$\Delta=1.4$ in Fig. \ref{fig:kozos1}. For $\Delta>1$ the function
$\tilde F_B$ always starts with a negative slope (corresponding to
\eqref{massiveT0}). We observed that as we lower $\Delta$ the position
of the local maximum stays approximately constant, but the local
minimum is shifted towards smaller temperatures.
We also plotted the curve for $\Delta=0.7$ and found that it is
increasing for small temperatures and has only a local maximum. We
attempted to compare the data at $\Delta=0.7$ to the QFT predictions,
but we found a mismatch. Our current numerical programs could only reach the temperature
regime of $T\approx 10^{-2}$ and for this magnitude of $T$ we still observed
a linear growth in $\tilde F_B$. It is possible that for much smaller temperatures
the QFT result  \eqref{masslessQFT} becomes correct, and 
further numerical and/or analytical
investigations are needed to check or disprove this.

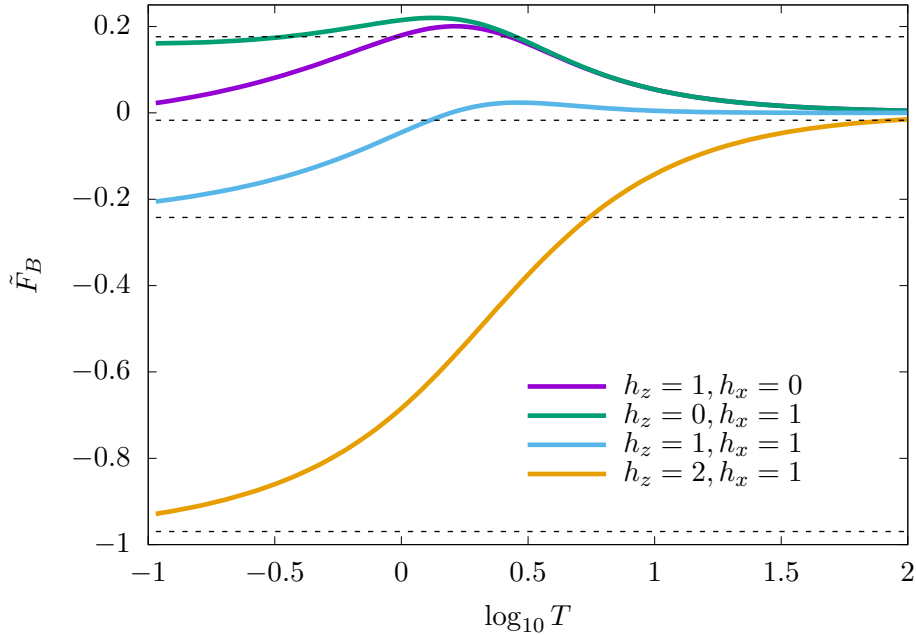
\begin{figure}
  \centering
  % AbraKozos2.plt
\input{AbraKozos2}
  \caption{The boundary free energy at anisotropy $\Delta=1.4$ as a function of the temperature,
    for different values of the boundary magnetic fields. We plotted $\tilde F_B=F_B-\Delta/2$
    such that all curves tend to 0 in the high temperature limit. The
    horizontal dashed lines show the $T\to 0$ limit as calculated from
    \eqref{DE12}-\eqref{eab}.}
  \label{fig:kozos2}
\end{figure}

We also evaluated $\tilde F_B$ in the cases with finite boundary fields.
Examples of the data can be seen in Fig. \ref{fig:kozos2},
where we plot $\tilde F_B$ at $\Delta=1.4$ for 
different values of $h_z$ and $h_x$. At intermediate and high
temperatures $\tilde F_B$ can be both increasing and decreasing,
depending on the magnitude of the boundary fields; this is in
accordance with the $\ordo(T^{-1})$ terms in the high temperature
expansion, as given by \eqref{highT2}.  At low temperatures all curves with $h_z\ne 0$ are increasing, as
given by the result \eqref{massiveT0}. 
The curve with $h_z=0$ and $h_x=1$ seems to be increasing at low $T$
in the available temperature range, although 
formula \eqref{massiveT0} gives a negative $\ordo(T)$ term. This curve
probably has a local minimum at some smaller value of the temperature
which is not accessible to our present programs; our data with
$h_z=h_x=0$ also showed that the local minima can be shifted to
surprisingly small values of $T$.

\section{Conclusions and discussion}

\label{sec:conclusions}

In this paper we treated the open XXZ spin chain and derived exact
results for the boundary free energy and boundary magnetization, with
the final formulas being \eqref{final} and \eqref{finalmag}. Special
cases of the general result were presented in subsection
\ref{ref:special}. In this section we discuss our results and the
open problems.

One of the key elements of our computation is the conjectured overlap
formula \eqref{OVERLAPS3}. In the diagonal case (longitudinal boundary
fields) it can be proven by known methods, but for generic
$K$-matrices we could only check it numerically
at finite Trotter number. Its validity is also corroborated by the fact
that the final results for $F_B$ agree with exact diagonalization and
DMRG and 
they produce the correct high and low temperature
limits. Nevertheless, an actual proof at finite Trotter number is desirable.

The overlap formula assumes that the Bethe roots
of the eigenstates of the QTM display the pair structure. It was shown
in \cite{sajat-minden-overlaps} that
 the pair structure for the non-vanishing overlaps follows from the
 Boundary Yang-Baxter relations. The proof of
 \cite{sajat-minden-overlaps} was given for the case of the physical
 spin chain, but it can be extended readily to the inhomogeneous case
 of the QTM. However, there is a limitation: the bulk magnetic field
 $h$ of the physical chain has to be zero. A finite bulk field would lead to twisted boundary
 conditions for the QTM \cite{kluemper-QTM,kluemper-review} which
 would break the pair structure of the Bethe roots.
The bulk magnetic
field is only compatible with the diagonal $K$-matrices, which
preserve the $U(1)$ symmetry of the model. Therefore, exact results
for the overlaps with $h\ne 0$ can only exist in the diagonal
case.
 Numerical 
 calculations show that for $h\ne 0$ more Bethe states have non-zero
 overlaps, and
 not only those that are the continuous deformations of 
 the Bethe states with the pair structure. Therefore we conclude that for
 $h\ne 0$ the structure of the overlaps is drastically different, and
 we can not expect a simple modification of \eqref{OVERLAPS3}.
We remind that the exact overlaps $h\ne 0$ were actually derived
already in \cite{sajat-karol}, but the final results for $F_B$ were more cumbersome than our
present formulas. In fact, the goal of deriving the Gaudin-like 
determinants was the primary motivation for this work.
  
In subsection \ref{sec:TBA} we explained that $F_B$ could be derived
alternatively in the framework of TBA, but the resulting expressions
would be considerably more complicated. For example, the Fredholm
operators would act also on the string indices. Nevertheless it is
desirable to pursue this approach as well. For the periodic system the equivalence of the QTM
and the TBA is understood \cite{TBA-QTM-Kluemper-Takahashi}, 
and it is desirable to establish the
relation also in the boundary problem.
An advantage of the TBA approach is that it could provide $F_B$ also 
in the case of a finite bulk magnetic field, because it only uses
statistical arguments \cite{sajat-g,woynarovich-uj} and the magnetic field is only a parameter of
the source term of the TBA.
Combined with our present
formulas this could lead to manageable expressions even at finite
$h$. 

It is an important task to compute the low-temperature behaviour of
$F_B$ in the massless regime, which could confirm the field
theoretical predictions. In \ref{sec:masslessT0} we computed the
boundary energy,
but the computation of the
first non-trivial correction is more involved. For the temperatures accessible to our present
numerical programs ($T\approx 10^{-2}-10^{-3}$) the boundary free energy
seemed to have a linear growth, which contradicts the predictions.
If the QFT result is still correct, it only
describes $F_B$ on a much smaller temperature scale than expected. 
From an analytical point of view the treatment of the isotropic point would be especially
interesting, because in this case $F_B$ is known to have logarithmic
terms in $T$ \cite{Affleck-XXZ-boundary-QFT}. The 
asymptotic analysis of the XXX case could be performed starting with formula \eqref{finalxxx}.

In Section \ref{sec:numerics} we demonstrated the numerical evaluation
of the exact results. We have found a wide variety of
behaviour as a function of the anisotropy, the boundary fields, and
the temperature. Perhaps surprisingly it was found that $F_B$ can be a
non-monotonic function of the temperature, and the sign of its
derivative  depends on all parameters. This is already
reflected by the high temperature
expansion \eqref{highT2}, which shows that the $\ordo(1/T)$ terms can be
positive or negative as well.
So far we only concentrated on the numerics for $F_B$, but it would be
desirable to compute the boundary magnetization, and the boundary
contributions to the entropy and the specific heat as well. These
quantities are given by different derivatives of $F_B$. The
exploration of these quantities is left to further research.

Finally we comment on the extension of these results to
other models. In those cases when there is a QTM formulation of the
thermodynamics and the boundary $K$-matrices are also known, our
methods could be applied; a crucial step would be to find the
generalization of the overlap formula \eqref{OVERLAPS3} using the 
techniques of \cite{sajat-minden-overlaps}. An example could be the
spin-1 XXZ chain, where both the QTM \cite{junji-higher-spin-qtm} and
the Boundary-ABA \cite{kulish-resh-sklyanin--fusion,zhou-fused-k} are
already worked out. 
We should note that there have been recent works dealing with impurity
thermodynamics within the QTM framework
\cite{klumper-talk,andrei-kondo-uj}. Possible relations to our results
need to be explored by future works.

\vspace{1cm}
{\bf Acknowledgments} 

We would like to thank to Frank G\"ohmann, Jesko Sirker, Yunfeng Yiang
and G\'abor Tak\'acs for useful discussions about this work. In
particular we are grateful to Jesko Sirker for sending us unpublished
DMRG data. We
 acknowledge support from the ``Premium'' Postdoctoral
Program of the Hungarian Academy of Sciences, the K2016 grant
no. 119204 and the KH-17 grant no. 125567 of the research agency
NKFIH.
This research was also supported by the Higher Education Institutional
Excellence Program announced by the Ministry of Human Resources, under
the thematic program Nanotechnology and Materials Science of the
Budapest University of Technology and Economics.

\bigskip

\addcontentsline{toc}{section}{References}
\providecommand{\href}[2]{#2}\begingroup\raggedright\endgroup

\end{document}

%% file: AbraKozos.tex
% GNUPLOT: LaTeX picture with Postscript
\begingroup
  \makeatletter
  \providecommand\color[2][]{%
    \GenericError{(gnuplot) \space\space\space\@spaces}{%
      Package color not loaded in conjunction with
      terminal option `colourtext'%
    }{See the gnuplot documentation for explanation.%
    }{Either use 'blacktext' in gnuplot or load the package
      color.sty in LaTeX.}%
    \renewcommand\color[2][]{}%
  }%
  \providecommand\includegraphics[2][]{%
    \GenericError{(gnuplot) \space\space\space\@spaces}{%
      Package graphicx or graphics not loaded%
    }{See the gnuplot documentation for explanation.%
    }{The gnuplot epslatex terminal needs graphicx.sty or graphics.sty.}%
    \renewcommand\includegraphics[2][]{}%
  }%
  \providecommand\rotatebox[2]{#2}%
  \@ifundefined{ifGPcolor}{%
    \newif\ifGPcolor
    \GPcolorfalse
  }{}%
  \@ifundefined{ifGPblacktext}{%
    \newif\ifGPblacktext
    \GPblacktexttrue
  }{}%
  % define a \g@addto@macro without @ in the name:
  \let\gplgaddtomacro\g@addto@macro
  % define empty templates for all commands taking text:
  \gdef\gplbacktext{}%
  \gdef\gplfronttext{}%
  \makeatother
  \ifGPblacktext
    % no textcolor at all
    \def\colorrgb#1{}%
    \def\colorgray#1{}%
  \else
    % gray or color?
    \ifGPcolor
      \def\colorrgb#1{\color[rgb]{#1}}%
      \def\colorgray#1{\color[gray]{#1}}%
      \expandafter\def\csname LTw\endcsname{\color{white}}%
      \expandafter\def\csname LTb\endcsname{\color{black}}%
      \expandafter\def\csname LTa\endcsname{\color{black}}%
      \expandafter\def\csname LT0\endcsname{\color[rgb]{1,0,0}}%
      \expandafter\def\csname LT1\endcsname{\color[rgb]{0,1,0}}%
      \expandafter\def\csname LT2\endcsname{\color[rgb]{0,0,1}}%
      \expandafter\def\csname LT3\endcsname{\color[rgb]{1,0,1}}%
      \expandafter\def\csname LT4\endcsname{\color[rgb]{0,1,1}}%
      \expandafter\def\csname LT5\endcsname{\color[rgb]{1,1,0}}%
      \expandafter\def\csname LT6\endcsname{\color[rgb]{0,0,0}}%
      \expandafter\def\csname LT7\endcsname{\color[rgb]{1,0.3,0}}%
      \expandafter\def\csname LT8\endcsname{\color[rgb]{0.5,0.5,0.5}}%
    \else
      % gray
      \def\colorrgb#1{\color{black}}%
      \def\colorgray#1{\color[gray]{#1}}%
      \expandafter\def\csname LTw\endcsname{\color{white}}%
      \expandafter\def\csname LTb\endcsname{\color{black}}%
      \expandafter\def\csname LTa\endcsname{\color{black}}%
      \expandafter\def\csname LT0\endcsname{\color{black}}%
      \expandafter\def\csname LT1\endcsname{\color{black}}%
      \expandafter\def\csname LT2\endcsname{\color{black}}%
      \expandafter\def\csname LT3\endcsname{\color{black}}%
      \expandafter\def\csname LT4\endcsname{\color{black}}%
      \expandafter\def\csname LT5\endcsname{\color{black}}%
      \expandafter\def\csname LT6\endcsname{\color{black}}%
      \expandafter\def\csname LT7\endcsname{\color{black}}%
      \expandafter\def\csname LT8\endcsname{\color{black}}%
    \fi
  \fi
    \setlength{\unitlength}{0.0500bp}%
    \ifx\gptboxheight\undefined%
      \newlength{\gptboxheight}%
      \newlength{\gptboxwidth}%
      \newsavebox{\gptboxtext}%
    \fi%
    \setlength{\fboxrule}{0.5pt}%
    \setlength{\fboxsep}{1pt}%
\begin{picture}(7200.00,5040.00)%
    \gplgaddtomacro\gplbacktext{%
      \csname LTb\endcsname%
      \put(814,704){\makebox(0,0)[r]{\strut{}$0$}}%
      \put(814,1383){\makebox(0,0)[r]{\strut{}$0.1$}}%
      \put(814,2061){\makebox(0,0)[r]{\strut{}$0.2$}}%
      \put(814,2740){\makebox(0,0)[r]{\strut{}$0.3$}}%
      \put(814,3418){\makebox(0,0)[r]{\strut{}$0.4$}}%
      \put(814,4097){\makebox(0,0)[r]{\strut{}$0.5$}}%
      \put(814,4775){\makebox(0,0)[r]{\strut{}$0.6$}}%
      \put(946,484){\makebox(0,0){\strut{}$-1.5$}}%
      \put(1783,484){\makebox(0,0){\strut{}$-1$}}%
      \put(2619,484){\makebox(0,0){\strut{}$-0.5$}}%
      \put(3456,484){\makebox(0,0){\strut{}$0$}}%
      \put(4293,484){\makebox(0,0){\strut{}$0.5$}}%
      \put(5130,484){\makebox(0,0){\strut{}$1$}}%
      \put(5966,484){\makebox(0,0){\strut{}$1.5$}}%
      \put(6803,484){\makebox(0,0){\strut{}$2$}}%
    }%
    \gplgaddtomacro\gplfronttext{%
      \csname LTb\endcsname%
      \put(176,2739){\rotatebox{-270}{\makebox(0,0){\strut{}$\tilde F_B$}}}%
      \put(3874,154){\makebox(0,0){\strut{}$\log_{10} T$}}%
      \csname LTb\endcsname%
      \put(2378,1951){\makebox(0,0)[l]{\strut{}$\Delta=1.6$}}%
      \csname LTb\endcsname%
      \put(2378,1731){\makebox(0,0)[l]{\strut{}$\Delta=1.4$}}%
      \csname LTb\endcsname%
      \put(2378,1511){\makebox(0,0)[l]{\strut{}$\Delta=0.7$}}%
    }%
    \gplbacktext
    \put(0,0){\includegraphics{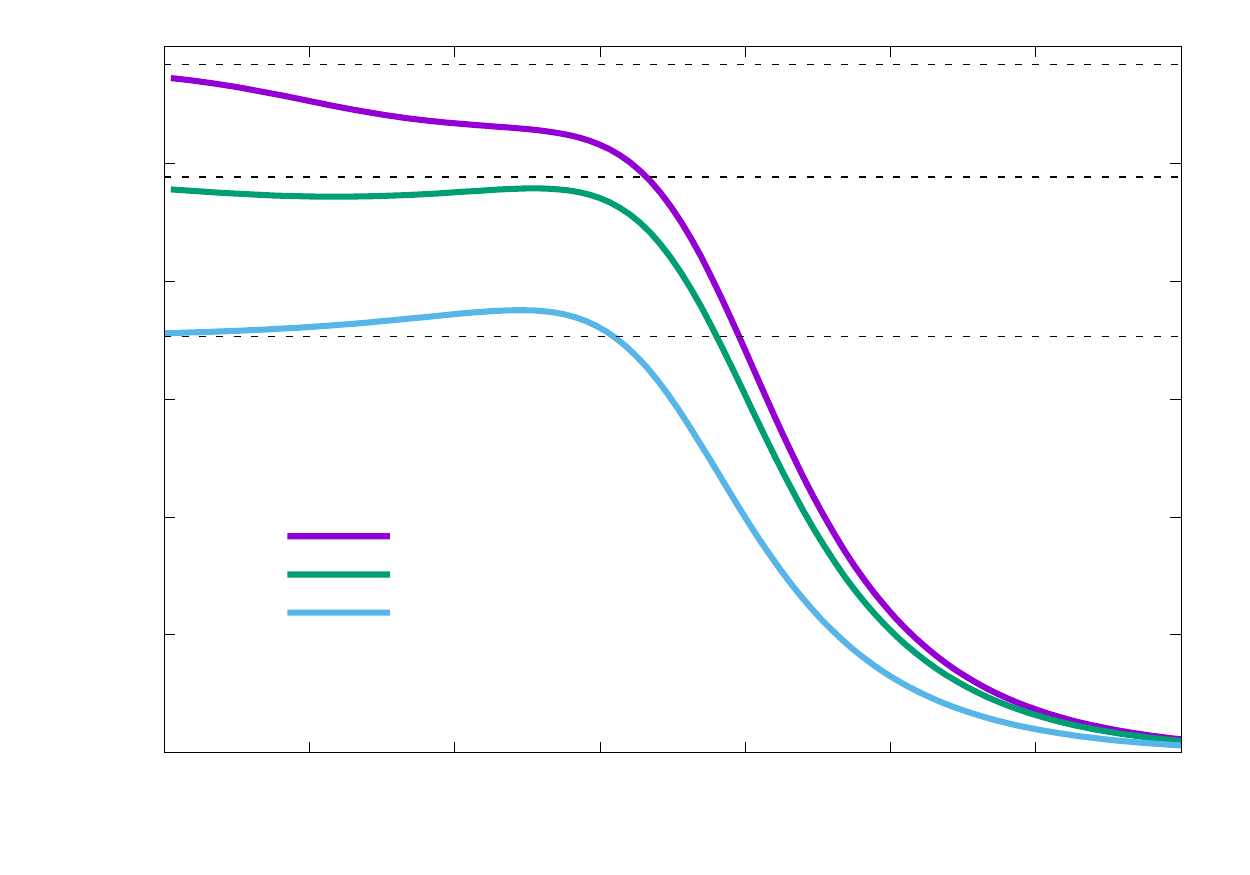}}%
    \gplfronttext
  \end{picture}%
\endgroup

%% file: AbraKozos2.tex
% GNUPLOT: LaTeX picture with Postscript
\begingroup
  \makeatletter
  \providecommand\color[2][]{%
    \GenericError{(gnuplot) \space\space\space\@spaces}{%
      Package color not loaded in conjunction with
      terminal option `colourtext'%
    }{See the gnuplot documentation for explanation.%
    }{Either use 'blacktext' in gnuplot or load the package
      color.sty in LaTeX.}%
    \renewcommand\color[2][]{}%
  }%
  \providecommand\includegraphics[2][]{%
    \GenericError{(gnuplot) \space\space\space\@spaces}{%
      Package graphicx or graphics not loaded%
    }{See the gnuplot documentation for explanation.%
    }{The gnuplot epslatex terminal needs graphicx.sty or graphics.sty.}%
    \renewcommand\includegraphics[2][]{}%
  }%
  \providecommand\rotatebox[2]{#2}%
  \@ifundefined{ifGPcolor}{%
    \newif\ifGPcolor
    \GPcolorfalse
  }{}%
  \@ifundefined{ifGPblacktext}{%
    \newif\ifGPblacktext
    \GPblacktexttrue
  }{}%
  % define a \g@addto@macro without @ in the name:
  \let\gplgaddtomacro\g@addto@macro
  % define empty templates for all commands taking text:
  \gdef\gplbacktext{}%
  \gdef\gplfronttext{}%
  \makeatother
  \ifGPblacktext
    % no textcolor at all
    \def\colorrgb#1{}%
    \def\colorgray#1{}%
  \else
    % gray or color?
    \ifGPcolor
      \def\colorrgb#1{\color[rgb]{#1}}%
      \def\colorgray#1{\color[gray]{#1}}%
      \expandafter\def\csname LTw\endcsname{\color{white}}%
      \expandafter\def\csname LTb\endcsname{\color{black}}%
      \expandafter\def\csname LTa\endcsname{\color{black}}%
      \expandafter\def\csname LT0\endcsname{\color[rgb]{1,0,0}}%
      \expandafter\def\csname LT1\endcsname{\color[rgb]{0,1,0}}%
      \expandafter\def\csname LT2\endcsname{\color[rgb]{0,0,1}}%
      \expandafter\def\csname LT3\endcsname{\color[rgb]{1,0,1}}%
      \expandafter\def\csname LT4\endcsname{\color[rgb]{0,1,1}}%
      \expandafter\def\csname LT5\endcsname{\color[rgb]{1,1,0}}%
      \expandafter\def\csname LT6\endcsname{\color[rgb]{0,0,0}}%
      \expandafter\def\csname LT7\endcsname{\color[rgb]{1,0.3,0}}%
      \expandafter\def\csname LT8\endcsname{\color[rgb]{0.5,0.5,0.5}}%
    \else
      % gray
      \def\colorrgb#1{\color{black}}%
      \def\colorgray#1{\color[gray]{#1}}%
      \expandafter\def\csname LTw\endcsname{\color{white}}%
      \expandafter\def\csname LTb\endcsname{\color{black}}%
      \expandafter\def\csname LTa\endcsname{\color{black}}%
      \expandafter\def\csname LT0\endcsname{\color{black}}%
      \expandafter\def\csname LT1\endcsname{\color{black}}%
      \expandafter\def\csname LT2\endcsname{\color{black}}%
      \expandafter\def\csname LT3\endcsname{\color{black}}%
      \expandafter\def\csname LT4\endcsname{\color{black}}%
      \expandafter\def\csname LT5\endcsname{\color{black}}%
      \expandafter\def\csname LT6\endcsname{\color{black}}%
      \expandafter\def\csname LT7\endcsname{\color{black}}%
      \expandafter\def\csname LT8\endcsname{\color{black}}%
    \fi
  \fi
    \setlength{\unitlength}{0.0500bp}%
    \ifx\gptboxheight\undefined%
      \newlength{\gptboxheight}%
      \newlength{\gptboxwidth}%
      \newsavebox{\gptboxtext}%
    \fi%
    \setlength{\fboxrule}{0.5pt}%
    \setlength{\fboxsep}{1pt}%
\begin{picture}(7200.00,5040.00)%
    \gplgaddtomacro\gplbacktext{%
      \csname LTb\endcsname%
      \put(946,704){\makebox(0,0)[r]{\strut{}$-1$}}%
      \put(946,1355){\makebox(0,0)[r]{\strut{}$-0.8$}}%
      \put(946,2007){\makebox(0,0)[r]{\strut{}$-0.6$}}%
      \put(946,2658){\makebox(0,0)[r]{\strut{}$-0.4$}}%
      \put(946,3309){\makebox(0,0)[r]{\strut{}$-0.2$}}%
      \put(946,3961){\makebox(0,0)[r]{\strut{}$0$}}%
      \put(946,4612){\makebox(0,0)[r]{\strut{}$0.2$}}%
      \put(1078,484){\makebox(0,0){\strut{}$-1$}}%
      \put(2032,484){\makebox(0,0){\strut{}$-0.5$}}%
      \put(2986,484){\makebox(0,0){\strut{}$0$}}%
      \put(3941,484){\makebox(0,0){\strut{}$0.5$}}%
      \put(4895,484){\makebox(0,0){\strut{}$1$}}%
      \put(5849,484){\makebox(0,0){\strut{}$1.5$}}%
      \put(6803,484){\makebox(0,0){\strut{}$2$}}%
    }%
    \gplgaddtomacro\gplfronttext{%
      \csname LTb\endcsname%
      \put(176,2739){\rotatebox{-270}{\makebox(0,0){\strut{}$\tilde F_B$}}}%
      \put(3940,154){\makebox(0,0){\strut{}$\log_{10} T$}}%
      \csname LTb\endcsname%
      \put(4661,1897){\makebox(0,0)[l]{\strut{}$h_z=1,h_x=0$}}%
      \csname LTb\endcsname%
      \put(4661,1677){\makebox(0,0)[l]{\strut{}$h_z=0,h_x=1$}}%
      \csname LTb\endcsname%
      \put(4661,1457){\makebox(0,0)[l]{\strut{}$h_z=1,h_x=1$}}%
      \csname LTb\endcsname%
      \put(4661,1237){\makebox(0,0)[l]{\strut{}$h_z=2,h_x=1$}}%
    }%
    \gplbacktext
    \put(0,0){\includegraphics{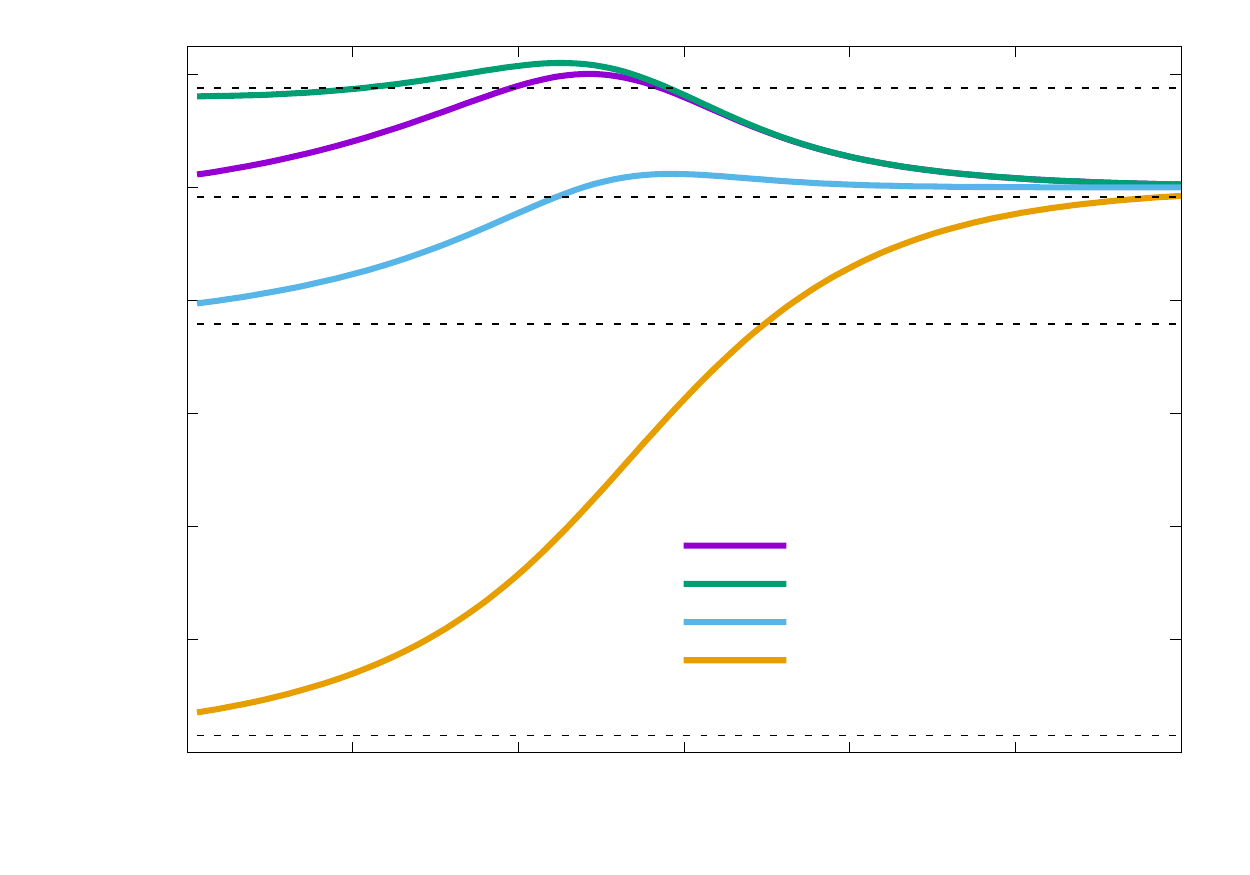}}%
    \gplfronttext
  \end{picture}%
\endgroup